\documentclass[10pt]{article}
\usepackage[utf8]{inputenc}
\usepackage{times}
\usepackage{graphicx}
\usepackage{amsmath}
\usepackage{amssymb}
\usepackage{authblk}
\usepackage{xcolor}
\usepackage{cite}

\usepackage{lineno}
\usepackage{comment}
\usepackage{multirow}
\mathchardef\mhyphen="2D % Define a "math hyphen"

\usepackage[a4paper,top=3cm,bottom=2cm,left=3cm,right=3cm,marginparwidth=1.75cm]{geometry}

\title{Stability of gene regulatory networks}
\author[1,2]{Yipei Guo}
\author[1]{Ariel Amir\thanks{Corresponding author, email: arielamir@seas.harvard.edu}}
\affil[1]{John A. Paulson School of Engineering and Applied Sciences, Harvard University, Cambridge, MA 02138}
\affil[2]{Program in Biophysics, Harvard University, Boston, MA 02115}

\date{}
\begin{document}

\maketitle
\begin{abstract}
    Homeostasis of protein concentrations in cells is crucial for their proper functioning, requiring steady-state concentrations to be stable to fluctuations.  Since gene expression is regulated by proteins such as transcription factors (TFs), the full set of proteins within the cell constitutes a large system of interacting components, which can become unstable. We explore factors affecting stability by coupling the dynamics of mRNAs and proteins in a growing cell. We find that mRNA degradation rate does not affect stability, contrary to previous claims. However, global structural features of the network can dramatically enhance
stability. Importantly, a network resembling a bipartite graph with a lower fraction of interactions that target TFs has a higher chance of being stable. Scrambling the $\textit{E. coli}$ transcription network, we find that the biological network is significantly more stable than its randomized counterpart, suggesting that stability constraints may have
shaped network structure during the course of evolution.          
\end{abstract}
%other global structural features of the networks (e.g., resembling a bipartite graph with a low fraction of interactions that target TFs) also affect the stability of the system. By scrambling the $\textit{E. coli.}$ transcription network, we show that the topology of such real networks can stabilize the system when the randomized network with the same number of regulatory interactions is unstable. These findings suggest that constraints imposed by system stability played a role in shaping the existing regulatory network during the evolutionary process. 

% We also find that in addition to the number of proteins and the density and strength of the regulatory interactions (as in May's approach), other global structural features of the networks (e.g., resembling a bipartite graph with a low fraction of interactions that target TFs) can also affect the stability of the system. 

\section*{Introduction}
Cells require different protein levels to survive in different external environments. The expression of these proteins within the cell are therefore highly regulated. An important regulatory mechanism involves transcription factors (TFs), which are themselves proteins that can either up or down regulate the transcription of mRNAs coding for other proteins by binding to enhancer or promoter regions of the regulated gene \cite{alon2019introduction}. Despite the importance of maintaining desired protein concentrations within cells, factors affecting the stability of these concentrations to perturbations have received little attention.

One approach of studying the stability of such systems with a large number of interacting components was introduced by May in the 1970s in the context of complex ecological communities \cite{may1972will}. The idea is that in a $n$-species community, the dynamics of the abundances $N_i$ of each species may in general be described by a set of ordinary differential equations:
\begin{equation}
    \frac{dN_i}{dt} = f_i(N_1,N_2,...N_n) 
    \label{eqn:dNidt}
\end{equation}
 for $i=1,2,...,n$, with corresponding steady-state solution ${N_i^{ss}}$ such that $f_i({\vec{N}^{ss}})=0$ $\forall$ $i$. The dynamics of small perturbations about this steady-state $x_i(t) = N_i(t)-N_i^{ss}$, when linearized about $N_i^{ss}$, has the form:
 \begin{equation}
     \frac{d\vec{x}}{dt} = \textbf{A} \vec{x},
 \end{equation}
 where $\textbf{A}$ is the Jacobian matrix with elements $A_{ij} = \left(\frac{\partial f_i}{\partial N_j} \right)^{ss}$. If all the eigenvalues of $\textbf{A}$ have a negative real part, the system relaxes back to the steady-state upon perturbations and the steady-state is said to be stable; if any of the eigenvalues have a positive real part, the steady-state is unstable as the system will move away from it (exponentially fast) when infinitesimally perturbed. To construct $\textbf{A}$, one would need to precisely know the functions $f_i$, which is often hard to obtain. May's approach was to model $\textbf{A}$ as a random matrix with independent, identically distributed off-diagonal elements (with mean $0$, standard deviation $\sigma$, and fraction of non-zero elements $C$) and constant diagonal elements $-a$. In the context of ecology, $\sigma$ reflects the average interaction strength between species, $C$ is the density of interactions or the probability that any two species interact, while $a$ is the self-regulation term which sets the relaxation time-scale of the system if there were no other pairwise interactions. From random matrix theory (RMT) and in particular the circular law for matrix eigenvalue distributions \cite{ginibre1965statistical, girko1985circular}, this system is stable if and only if $a > \sigma \sqrt{nC}$. This implies that the system becomes unstable above some critical size, and that increasing $a$ stabilizes the system and allows for stronger interactions between species. 

This approach has also been used to analyze other large interacting systems. In particular it has been used to argue why weak repressions by microRNAs, thought of as effectively increasing the degradation rate of mRNAs, confer stability to gene regulatory networks \cite{chen2019gene, zhao2017weak}. However, such a framework does not take into account the functional form of $f_i$ and in particular that the matrix elements often depend on the steady-state solutions themselves. These details of the model can be important $\--$ for example, when competition for resources between ecological species are explicitly modeled (using the MacArthur's consumer resource model), even when the interactions (i.e. preferences of each species for the different resources) are completely random, the spectrum of the Jacobian matrix that represents effective pairwise interaction between species is no longer circular (but rather, follows the Marchenko-Pastur distribution) \cite{cui2019diverse}. Furthermore, transcriptional regulatory networks are not random but instead have distinct structural features. The structure of interaction networks has been known to affect stability in other models \cite{allesina2012stability, thebault2010stability, okuyama2008network, cui2019diverse}. However, how these features affect the stability of gene regulatory networks has not been explored.
% For example, within the MacArthur's consumer resource model in ecology where interactions between species occur through competition for resources, it was found that including a non-random component into the structure of the resource consumption matrix (indicating the preference of each species for each resource) can stabilize the system \cite{cui2019diverse}. 

%which can give rise to correlations between elements and other structural features in the Jacobian matrix $\textbf{A}$.

Here, by analyzing a model that takes into account the transcription of mRNAs from genes, translation of mRNAs into proteins, and transcriptional regulation by proteins, we investigate the stability of this large system of coupled mRNAs and proteins in growing cells, and find that while the mRNA degradation rate can affect relaxation rate back to steady-state levels, it does not affect whether the system is stable. Instead, stability can depend strongly on global structural features of the interaction network. In particular, given the same number of proteins, TFs, number of interactions, and regulation strengths, a network with a lower fraction of interactions that target TFs has a higher chance of being stable. In the limit where there are no TF-TF interactions i.e. all TFs regulate proteins that are not TFs, it is possible for the system to remain stable for arbitrarily large system sizes, unlike random networks which become unstable when system size becomes too large. By scrambling the $\textit{E. coli.}$ transcription network, we find that the topology of real networks can stabilize the system since the randomized network with the same number of regulatory interactions is often unstable. These findings suggest that constraints imposed by system stability may have played a significant role in shaping the existing regulatory network during the evolutionary process. By carrying out the analysis for different physiological states the cell can be in (corresponding to different sets of dynamical equations) and with different choices of parameter distributions, we also show that our main results and conclusions are robust to the details of the model.

% We also show that while the coupling between mRNA and proteins (through the process of transcription, translation, and regulation) our main results  

% Here, by analyzing a model that takes into account the transcription of mRNAs from genes, translation of mRNAs into proteins, and transcriptional regulation by proteins, we investigate the stability of this large system of coupled mRNAs and proteins in growing cells, and find that while the mRNA degradation rate can affect relaxation rate back to steady-state levels, it does not affect whether the system is stable. Importantly, we also find that in addition to the number of proteins and the density and strength of the regulatory interactions (as in May's approach), other global structural features of the networks (e.g., resembling a bipartite graph with a low fraction of interactions that target TFs) also affect the stability of the system. By scrambling the $\textit{E. coli.}$ transcription network, we show that the topology of such real networks can stabilize the system when the randomized network with the same number of regulatory interactions is unstable. These findings suggest that constraints imposed by system stability played a role in shaping the existing regulatory network during the evolutionary process.

% These suggest the existence of stability constraints during the evolutionary process that gave rise to these networks.

\section*{Results}

\subsection*{The model}
\begin{figure}[ht]
    \centering
	\includegraphics[width=13.462cm]{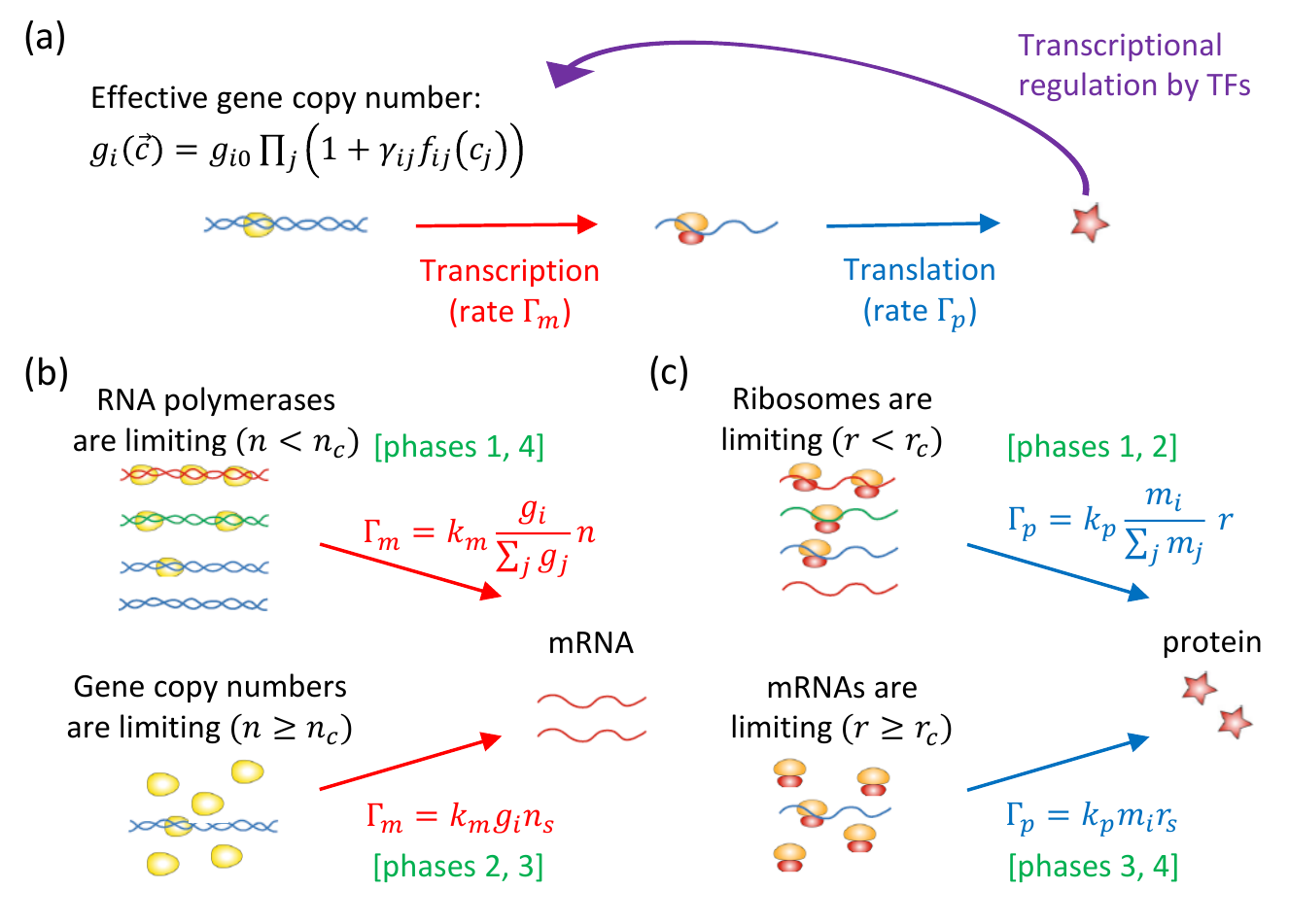}
    \caption{Schematic illustration of the gene expression model. (a) The dynamics of protein and mRNA concentrations are coupled through transcriptional regulation, where some of the proteins (e.g. transcription factors) modulate the effective gene copy numbers $g_i$ and hence the transcription rate of other genes. (b) If RNA polymerase is in excess, transcription rate $\Gamma_m$ of gene $i$ is proportional to its effective gene copy number $g_i$. If instead RNA polymerase is limiting, $\Gamma_m$ is proportional to the gene allocation fraction $\phi_i = g_i/\sum_j g_j$. (c) Translation rate $\Gamma_p$ is proportional to mRNA number $m_i$ if mRNAs are limiting, and proportional to the mRNA fraction $m_i/\sum_j m_j$ if ribosomes are limiting. There are 4 different phases of the model depending on whether RNA polymerases and ribosomes are limiting.}
    \label{fig:model}
\end{figure}

Gene expression involves two major steps: transcription and translation (Fig.\ref{fig:model}a). Transcription is the process in which mRNA is synthesized by RNA polymerase using DNA as a template. The transcription rate of a gene $i$ therefore depends on the number of RNA polymerases $n$ and its effective gene copy number $g_i$ which takes into account both its copy number and how strongly RNA polymerase can bind to the promoter of that gene \cite{lin2018homeostasis}. Due to the presence of TFs, $g_i(\vec{c})$ can in general depend on the set of protein concentrations $\vec{c}$ (Fig.\ref{fig:model}a). We assume that multiple TFs acting on the same gene act independently, with their effects stacking multiplicatively. This allows for both "OR"- and "AND"-gate-like combinatorial effects \cite{buchler2003schemes}, and can emerge from a thermodynamic model of TF binding (SI Section \ref{sec:ThermoModel}). Therefore, we adopt the following form for
transcriptional regulation throughout the paper:  
\begin{equation}
g_i(\vec{c}) = g_{i0} \prod_j (1 + \gamma_{ij} f_{ij}(c_j)),
\label{eqn:giexpression}
\end{equation}
where $g_{i0}$ is the effective gene copy number of $i$ if it were unregulated (randomly drawn from a uniform distribution), and $\gamma_{ij}$ controls the type and strength of regulation i.e. how much gene expression of $i$ changes in the presence of the TF $j$. In particular, $\gamma_{ij} > 0$ if $j$ up-regulates $i$ and $-1 \leq \gamma_{ij} < 0$ if $j$ down-regulates $i$. For each regulatory interaction, we assume that the fold-change $\Omega_{ij}$ is drawn from a uniform distribution between 1 and $\Omega_{max}$, such that 
\begin{align}
    \gamma_{ij} = \begin{cases}
    \Omega_{ij} - 1 & \text{if $\gamma_{ij} > 0$ (up-regulating)}
    \\
    \frac{1}{\Omega_{ij}} - 1 & \text{if $\gamma_{ij} < 0$ (down-regulating)}
    \end{cases}
    \label{eqn:gammadef}
\end{align}
since this would allow $g_i(c_j)$ to increase (if $j$ up-regulates $i$) or decrease (if $j$ down-regulates $i$) by a factor of $\Omega_{ij}$ in the limit of high $c_j$. In SI section \ref{sec:FCdistEffect}, we show that the main results do not depend on the particular distribution $P(\Omega)$ used. 

Motivated by experimental measurements of the relationship between transcription factor input and gene expression output showing a sigmoidal functional form of $f_{ij}(c_j)$ \cite{kuhlman2007combinatorial, kim2008quantitative}, we take it to be a Hill function
\begin{equation}
    f_{ij}(c_j) =  \frac{c_j^{n_{ij}}}{K_{ij}^{n_{ij}}+c_j^{n_{ij}}},
    \label{eqn:fijexpression}
\end{equation}
with $n_{ij}>0$. 

Following Ref. \cite{lin2018homeostasis}, we assume a threshold number $n_c$ of RNA polymerases above which the gene copy number is limiting the transcription rate (Fig.\ref{fig:model}b). When this is the case, transcription rate is proportional to $g_i$ and is independent of $n$. If instead $n<n_c$, it is the RNA polymerases that are limiting, in which case the different genes have to compete for the limited pool of RNA polymerases. The transcription rate of a gene $i$ is then proportional to both $n$ and the fraction of RNA polymerases working on that gene, the gene allocation fraction:
\begin{equation}
    \phi_i(\vec{c})= \frac{g_i(\vec{c})}{\sum_j g_j(\vec{c})}.  \label{eqn:phidef}
\end{equation}

Denoting the number of different genes by $N$, the dynamics of mRNA $m_i$ for $i = 1,...N$ can therefore be described by the following equation:
\begin{equation}
\begin{split}
    \frac{dm_i}{dt} &= \begin{cases}
    k_m \phi_i(\vec{c})n - \frac{m_i}{\tau_m} & \text{ if $n<n_c$}
    \\
    k_m g_i(\vec{c})n_s - \frac{m_i}{\tau_m} & \text{ if $n \geq n_c$}
    \end{cases}
    \label{eqn:dmdt}
\end{split}
\end{equation}
where $k_m$ characterizes the transcription rate of a single RNA polymerase, $\tau_m$ is the mRNA lifetime, and $n_s$ is the maximum number of RNA polymerases per gene.

Similarly for the process of translation where ribosomes make proteins using mRNA as a template, the translation rate depends on the number of ribosomes $r$ and the mRNA copy number $m_i$. As for RNA polymerases, there is also a threshold number of ribosomes $r_c$ above which mRNA number is limiting and below which ribosomes are limiting (Fig.\ref{fig:model}c). The dynamics of protein numbers $p_i$ for $i = 1,..., N$, with $p_{N-1} = n$ corresponding to RNA polymerases and $p_N = r$ corresponding to ribosomes, are therefore given by:
\begin{equation}
\begin{split}
    \frac{dp_i}{dt} &= \begin{cases}
    k_p \frac{m_i}{\sum_j m_j} r - \frac{p_i}{\tau_p} & \text{ if $r<r_c$}
    \\
    k_p m_i r_s - \frac{p_i}{\tau_p} & \text{ if $r \geq r_c$}
    \end{cases}
    \label{eqn:dpdt}
\end{split}
\end{equation}
where $k_p$ characterizes the translation rate of a single ribosome, $\tau_p$ is the protein lifetime, and $r_s$ is the number of ribosomes per mRNA when ribosomes are in excess.
%, both of which can be regulated by TFs but cannot themselves be TFs

Depending on whether the RNA polymerases and ribosomes are limiting, there are 4 different cellular phases (Fig.\ref{fig:model}b, c). The regime where $n \geq n_c$ and $r \geq r_c$ (phase 3 of the model, where the production rate of mRNAs and proteins are proportional to gene and mRNA copy numbers respectively) has been widely studied \cite{paulsson2005models, shahrezaei2008analytical, thattai2001intrinsic}, but has been shown to be inconsistent with experimental observations in wild type cells showing exponential growth of protein levels \cite{zhurinsky2010coordinated, schmidt1995cell}. Instead, the regime where $n<n_c$ and $r<r_c$ (phase 1 of the model) is the one where wild type fission yeast \cite{zhurinsky2010coordinated} and mammalian cells appear to be in \cite{schmidt1995cell}. We therefore focus on this phase for the rest of the paper. Note, however, that the phase 3 regime has been experimentally observed in defective budding yeast and mammalian cells that are excessively large \cite{neurohr2019excessive}, whereas the regime where RNA polymerases are in excess ($n \geq n_c$) while ribosomes are limiting ($r<r_c$) (phase 2 of the model) has been observed in mutant fission yeast \cite{zhurinsky2010coordinated}. We will address these two phases in the SI. The regime where $n<n_c$ and $r \geq r_c$ (phase 4 of the model) is biologically unrealistic as ribosomes are typically more expensive to make compared to other proteins and hence having excess ribosomes while RNA polymerases are limited would be inefficient \cite{reuveni2017ribosomes, scott2010interdependence}. This regime is therefore not considered.
 
% When the production rate of proteins exceeds the degradation rate, the volume of the cell $V$ increases over time. It it therefore relevant to look at the corresponding 
It will be convenient to consider the dynamics for the \textit{concentrations} of mRNAs $c_{mi} = \frac{m_i}{V}$ and proteins $c_i = \frac{p_i}{V}$. In bacteria \cite{kubitschek1984independence, basan2015inflating} and mammalian cells \cite{crissman1973rapid}, the volume of the cell $V$ is approximately proportional to the total protein mass. Hence, we assume for simplicity that each protein has the same mass and set the cell density to be 1, such that $V = \sum_i p_i$. The dynamics for concentrations in phase 1 are then given by:
\begin{equation}
    \frac{dc_{mi}}{dt} = 
    k_m \phi_i(\vec{c}) c_n - c_{mi} \left(k_p c_r + \frac{1}{\tau}\right) 
    \label{eqn:dcmdt}
\end{equation}
\begin{equation}
    \frac{dc_i}{dt} = k_p c_r \left( \frac{c_{mi}}{c_{mT}} - c_i \right) 
    \label{eqn:dcpdt}
\end{equation}
where $c_{mT} = \sum_i c_{mi}$ is the total concentration of all mRNAs and $\frac{1}{\tau} = \frac{1}{\tau_m} - \frac{1}{\tau_p}$ is the difference between mRNA and protein degradation rates (which can be positive or negative). A summary of the list of model parameters can be found in the SI Section \ref{sec:ParameterList}, Table \ref{tab:params}. 

While this set of equations govern the dynamics of average concentrations and hence do not capture stochastic effects inherent in gene expression and in the binomial sampling of molecules during cell division, these fluctuations do not affect the average steady-state concentrations if the number of molecules is large (see SI Section \ref{sec:Gillespie}, Fig. \ref{SIfig:Gillespie}). In fact, these fluctuations can be considered as perturbations about steady-state values, and we investigate the stability of the system to such perturbations in the rest of the paper.

% In the limit where the lifetime of mRNAs is much shorter than the length of the cell cycle, which is typically true in wild-type cells, the relaxation dynamics of mRNA is much faster than that of proteins such that $\frac{dm_i}{dt} \approx 0$ at all times. The mRNA numbers are then given by:
% \begin{align}
%     \langle m_i^*(t) \rangle = \begin{cases}
%     k_m \phi_i \langle n(t) \rangle \tau_m & \text{in phase 1} 
%     \\
%     k_m g_i n_s \tau_m & \text{in phase 3}
%     \end{cases}
%     \label{eqn:mi_ss}
% \end{align}

% However, while the mRNA numbers are approximately constant throughout the cell cycle in phase 3 (Eqn.\ref{eqn:mi_ss}), in phase 1 it is the mRNA concentrations that are kept at constant steady state values given by
% \begin{equation}
%     c^*_{mi} = \frac{k_m c^*_n}{k_p c^*_r + \frac{1}{\tau}} \phi_i({c^*}) 
%     \label{eqn:cmi_ss}
% \end{equation}

% Consequently, another main difference between cells in phase 1 and phase 3 is that assuming that protein degradation rate is much slower than the protein production rates (which is typically true for growing wild-type cells), cell volume $V = \sum_i p_i$ grows exponentially with exponential growth rate 
% \begin{equation}
%     \mu_1 = k_p \phi_r - \frac{1}{\tau_p}
%     \label{eqn:mu1}
% \end{equation}
% in phase 1, while in phase 3 the cell grows linearly (Eqn.\ref{eqn:dpdt}). It is nevertheless convenient to define $\mu_3$ as the growth rate per unit volume in phase 3, which is given by 
% \begin{equation}
%     \mu_3 = \frac{k_m k_p r_s n_s \tau_m g_T}{V} - \frac{1}{\tau_p}.
%     \label{eqn:mu3}
% \end{equation}

\subsection*{Effects of network features and topology on stability of the system}
To study how properties of the transcriptional regulatory network affect the stability of the system, we first consider the regime where the lifetime of mRNAs is much shorter than that of proteins, which is typically true for wild-type cells \cite{milo2015cell}. In this limit of fast mRNA degradation, the relaxation dynamics of mRNA is much faster than that of proteins such that $\frac{dc_{mi}}{dt} \approx 0$ at all times. Eliminating the fast process (by substituting the steady-state mRNA concentrations $ c_{mi} = \frac{k_m c_{n}}{k_p c_{r} + \frac{1}{\tau}} \phi_i(\vec{c}) $ obtained from Eqn. \ref{eqn:dcmdt} into 
% From Eqn. \ref{eqn:dcmdt} we find that the mRNA concentrations at any time are then given by:
% \begin{equation}
%     c_{mi} = \frac{k_m c_{n}}{k_p c_{r} + \frac{1}{\tau}} \phi_i(\vec{c}).
%     \label{eqn:cmi_ss}
% \end{equation}
Eqn.\ref{eqn:dcpdt}), the dynamics of protein concentrations can be written as a set of $N$ ODEs:
\begin{align}
    \frac{dc_i}{dt}  \approx k_p c_r \left( \phi_i(\vec{c}) - c_i \right). 
    \label{eqn:dcpdt_smalltaum}
\end{align}
The stability of the system therefore depends only on the eigenvalues of the $N \times N$ Jacobian matrix $\mathbf{A} = k_p c_r^{ss} (\mathbf{M}-\mathbf{I})$, where we define the interaction matrix
\begin{equation}
 M_{ij} = \frac{\partial \phi_i}{\partial c_j}\vert_{\vec{c}={\vec{c}^{ss}}},
 \label{eqn:Mmatrix}
\end{equation}
with the steady-state protein concentrations given by $c_i^{ss} = \phi_i(\vec{c}^{ss})$ (from Eqn. \ref{eqn:dcpdt_smalltaum}).
% \begin{equation}
%     c_i^{ss} = \phi_i(\vec{c}^{ss}).
%     \label{eqn:c_ss}
% \end{equation}

Denoting $\lambda_M$ as the eigenvalues of $\textbf{M}$, the system is stable as long as the maximal real part of these eigenvalues $\lambda_{M,r_{max}}$ is smaller than $1$ (such that all eigenvalues of $\mathbf{A}$ have a negative real part). It is therefore useful to understand the structure of $\textbf{M}$ by breaking it into two parts using Eqn. \ref{eqn:phidef}:
\begin{equation}
    M_{ij} = c_i^{ss} (M_{1,ij} - M_{2,ij}),
    \label{eqn:Mdecomposed}
\end{equation}
where  
\begin{equation}
    M_{1,ij} = \frac{\partial \text{log} g_i}{\partial c_j}     
    \label{eqn:M1def}
\end{equation}
captures the direct interactions between proteins, while 
\begin{equation}
    M_{2,ij} = \frac{\partial \text{log}g_T}{\partial c_j} = \sum_k c_k^{ss} \frac{\partial \text{log} g_k}{\partial c_j} 
    \label{eqn:M2def}
\end{equation}
is a rank-1 matrix that captures the indirect interactions arising from competition for ribosomes. 

It can be shown that both the structure of $\bf{M}$ (Eqn.\ref{eqn:Mdecomposed}) and the fact that stability only depends on $\bf{M}$ still hold in the other phases, despite the exact equations for protein dynamics being different (see SI Section \ref{sec:differentphases}). Therefore, even though the simulations in the rest of this section are carried out in phase 1, our findings and conclusions also apply to the other phases.  

%This tendency for inhibitory regulations to destabilize the system can be intuitively understood as follows: In phase 1, a slight increase in regulator concentration from steady-state reduces the gene copy number and hence mRNA levels of the regulated gene. This in turn increase mRNA fraction (and hence protein concentrations) of the regulator to increase further. In phase 3, the reduction in mRNA levels of the regulated gene reduces the rate at which proteins are made. This slowing down of the increase in cell volume causes the regulator protein concentrations to increase.

\subsubsection*{Stability of the system scales with $\sqrt{N}$ for random regulatory networks.}

We start by exploring the stability of `fully random' regulatory networks, which we take to be our null model. 
%Based on RMT, if the elements of $\bf{M}$ are independently drawn from some distribution, one would expect its maximum eigenvalue to scale as $\sqrt{N}$. 

Since the maximum eigenvalue of a random matrix depends on the standard deviation of its elements, we first carry out a naive estimate of how the elements of $\bf{M}$ scale with $N$. With $g_i({c})$ given by Eqn. \ref{eqn:giexpression}, 
\begin{equation}
    \frac{\partial \text{log} g_i}{\partial c_j} = \frac{\gamma_{ij}}{1+\gamma_{ij} f_{ij}(c_j)} \frac{\partial f_{ij}}{\partial c_j}.
\end{equation}
Biologically, TF concentrations are often comparable to the values of dissociation constants $K_d$ for DNA binding \cite{milo2015cell}. Therefore, since $c_j \sim 1/N$, we also choose $K_{ij} \sim 1/N$ (Eqn.\ref{eqn:fijexpression}), which would allow cells to maintain the full range of gene expression response. From Eqn. \ref{eqn:fijexpression}, this implies that $f_{ij} \sim O(1)$ and $\frac{\partial f_{ij}}{\partial c_j} \sim N$, and hence $M_1$ and $M_2$ also scale with $N$ (Eqns. \ref{eqn:M1def}, \ref{eqn:M2def}). We therefore expect $M_{ij} \sim O(1)$ (Eqn. \ref{eqn:Mdecomposed}), and hence (from RMT), for $\lambda_{M,r_{max}}$ to scale approximately as $\sqrt{N}$ for random interaction networks. $\lambda_{M,r_{max}}$ also increases with the strength of the interactions $\gamma$, implying that the system will become unstable either when $N$ exceeds a critical number or the regulation strength becomes too high. However, this argument neglects correlations between the elements of $\bf{M}$, which could potentially be relevant. In fact, we will see in the later sections that the structure of $\bf{M}$ (Eqn. \ref{eqn:Mdecomposed}) plays an important role in influencing the stability of the system.

Therefore, to test if this scaling relation holds, we constructed networks of a specified interaction density $\rho$ by randomly selecting $\rho N^2$ interactions from the $N(N-1)$ possibilities (where we have assumed that ribosomes cannot act as TFs), and choose half of the interactions to be up-regulating with the remaining half being down-regulating. 
%namely $P(\Omega) \sim \frac{1}{\Omega}$ and $P(\Omega) \sim \frac{1}{\Omega^2}$, 

\begin{figure}[ht]
    \centering
	\includegraphics[width=8.9cm]{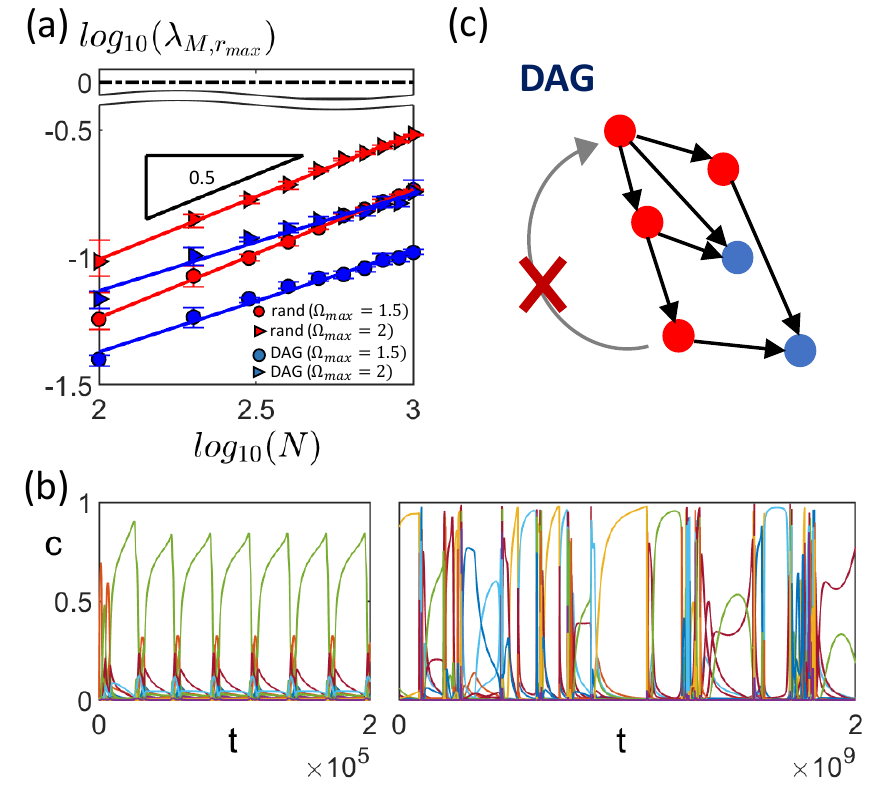}
    \caption{Stability of random interaction networks. (a) For random interaction networks (red markers,`rand'), the maximal real part of the eigenvalues of the interaction matrix $\lambda_{M,r_{max}}$ scales with $\sqrt{N}$. Surprisingly, for random directed acyclic networks (blue markers,`DAG'), $\lambda_{M,r_{max}}$ also scales approximately with $\sqrt{N}$. In both of these cases, increasing the interaction strength from $\Omega_{max}=1.5$ (circles) to $\Omega_{max}=2$ (triangles) increases $\lambda_{M,r_{max}}$. These results suggest that the system will become unstable (i.e. $log_{10}(\lambda_{M,r_{max}})$ exceeds 0, indicated by the black dashed line) when $N$ or $\Omega_{max}$ becomes too large. Each data point is obtained from an average of 10 randomly drawn networks, with error bars indicating the interquartile range. Each random network is constructed by randomly selecting $\rho N^2$ interactions from $N(N-1)$ possibilities, with half of the interactions chosen to be up-regulating and the remaining half to be down-regulating. Construction of DAGs is described in (c). For each regulatory interaction, fold change is chosen uniformly between $1$ and $\Omega_{max}$.  [Other parameters: $\rho = 0.01$, $n = 1$.] (b) When systems go out of stability, they exhibit oscillatory (left, $\Omega_{max} = 20$) followed by chaotic behavior (right, $\Omega_{max} = 200$) as interaction strengths are increased. [Other parameters: $N=200$, $\rho = 0.2$, $n = 1$, fully random network, time $t$ is in units of $1/k_p$.] (c) Random directed acyclic networks are constructed by randomly drawing connections between proteins (red circles represent TFs, blue circles represent non-TFs). If a drawn connection creates a loop (e.g. the grey arrow with a cross on it), it is rejected. }
    \label{fig:Stability_randomnetworks}
\end{figure}

By taking the ensemble average over the randomly drawn networks, we indeed recover the $\sqrt{N}$ scaling (Fig. \ref{fig:Stability_randomnetworks}a), which is also robust to the fraction of up- and down- regulatory interactions (see SI Section \ref{sec:IntSignEffect}, Fig. \ref{SIfig:EffectofIntSign}a) and the distribution of fold-changes $P(\Omega)$ (see SI Section \ref{sec:FCdistEffect}, Fig. \ref{SIfig:EffectofFCdist}). For sufficiently large $N$ or $\Omega_{max}$, we can no longer find the fixed point of the system. Nevertheless, by simulating the dynamics, we find that for interaction networks of a given $N$ and $\rho$, we get oscillatory, followed by chaotic behaviour as $\Omega_{max}$ is increased (Fig. \ref{fig:Stability_randomnetworks}b). Similar phenomena have also been described and analyzed in models of neural networks \cite{sompolinsky1988chaos} and ecological systems \cite{roy2019numerical}. While certain biochemical circuits have been known to generate oscillations such as in the cell cycle and the circadian clock, the oscillatory dynamics observed here is of a different nature $\--$ it does not come about from any specific fine-tuning of the network but, rather, emerges from having a large number of randomly and strongly interacting genes. 
% here are of a different nature, giving rise to oscillations in all of the proteins in the cell and  
% The oscillatory behavior observed here arises due to a loss in a stable fixed point as interaction strengths are increased. 

However, transcriptional regulatory networks are typically not random. Instead, they are enriched for distinct structural features such as the following motifs: feedforward loops (FFL), single input module (SIM) and dense overlapping regulons (DOR) which do not contain any loops besides autoregulatory ones \cite{alon2019introduction,shen2002network}. In the next few subsections we therefore explore the effects of network topology on the system stability.

\subsubsection*{Random directed acyclic networks can also be unstable.}
Since transcription networks as a whole resemble directed acyclic graphs (DAGs) \cite{shen2002network, alon2019introduction}, we explore the stability of such networks. 

In systems where the Jacobian matrix reflects the presence of direct interactions between components, the elements of the Jacobian matrix $A_{ij}$ is 0 if $j$ does not influence or regulate $i$. In such cases, if there are no interaction loops involving 2 or more components (e.g. E regulates F which also regulates E), $\mathbf{A}$ can be written as a triangular matrix for such a DAG and the eigenvalues are the diagonal elements of the matrix i.e. the self-regulation loops. The system is therefore stable if there are no auto-activation among the components i.e. there are no positive elements along the diagonal of $\mathbf{A}$.

In our case, the presence of indirect interactions captured by the additional $\mathbf{M_2}$ matrix (Eqn. \ref{eqn:Mdecomposed}) implies that even if the regulation network is a DAG, the stability of the system is not determined solely by the self-regulation loops. Instead, we find that if we draw DAGs randomly (constructed by adding a connection only if the resultant network is still acyclic, Fig.\ref{fig:Stability_randomnetworks}c), even if there are no self interactions, the largest eigenvalue still scales approximately with $\sqrt{N}$, suggesting that it is still possible for such a network to go unstable. Nevertheless, there is a negative offset in $\lambda_{M,r_{max}}$ compared to the fully random case (Fig. \ref{fig:Stability_randomnetworks}a), implying that the lack of loops does help to stabilize the system.

% \begin{figure}[ht]
%     \centering
% 	\includegraphics[width=8.7cm]{Fig4_DAGandBipartite_full_v2.pdf}
%     \caption{Effect of various network properties on stability of system. (a) Random directed acyclic networks are constructed by randomly drawing connections between proteins. If a drawn connection creates a loop (e.g. the grey arrow with a cross on it), it is rejected. (b) For random directed acyclic graphs, maximum eigenvalue of the Jacobian matrix $\lambda_{M,r_{max}}$ seem to scale as $\sqrt{N}$. (c) When constructing a bipartite interaction network, we group the proteins into TFs and non-TFs, and only allow directed regulatory interactions to go from a TF to a non-TF. (d) For bipartite regulatory networks, $\lambda_{M,r_{max}}$ is independent of $N$. Here, we assume the number of TFs $q = k N^2$, with $k = 5 \times 10^{-4}$. [Other parameters: $\rho = 0.01$.]}
%     \label{fig:DAGandBipartite}
% \end{figure}

\subsubsection*{Bipartite structure can maintain stability of large networks.}
A commonly found motif in the \textit{Escherichia coli} sensory transcription networks is the dense-overlapping regulons (DORs) which consist of a set of regulators that combinatorially control a set of output genes \cite{shen2002network, alon2019introduction, alon2007network}. There are several of these DORs in \textit{E. coli}, each with hundreds of output genes, and they appear to occur in a single layer i.e. there is no DOR at the output of another DOR. Such a structure can be thought of as a bipartite graph in which there are 2 types of nodes representing transcription factors (TFs) and non-transcription factors (non-TFs), and every directed edge go from a TF to a non-TF. Since such graphs do not contain any regulatory loops (and are therefore also DAGs), we expect them to be more stable than random networks. However, they are a specific subset of DAGs in which none of the TFs are themselves regulated. This is also a key difference between these networks and bipartite, mutualistic networks commonly studied in ecological models \cite{thebault2010stability, okuyama2008network}. In this subsection, we investigate the stability of such networks. 

To study this problem, we first group proteins into 2 categories: $q$ TFs and $N-q$ non-TFs, such that for any general network the components of the Jacobian matrix have the following structure:
\begin{equation}
    \mathbf{M_1} = \begin{pmatrix} 
    \mathbf{T_1} & \mathbf{0} \\
    \mathbf{R_1} & \mathbf{0} 
    \end{pmatrix}
    \label{eqn:M1decomposed}
\end{equation}
\begin{equation}
    \mathbf{M_2} = \begin{pmatrix} 
    \mathbf{T_2} & \mathbf{0} \\
    \mathbf{R_2} & \mathbf{0} 
    \end{pmatrix},
    \label{eqn:M2decomposed}
\end{equation}
where  $\mathbf{T_1}$ ($\mathbf{T_2}$) is a $q \times q$ matrix representing the direct (indirect) effect of TFs on TFs while $\mathbf{R_1}$ ($\mathbf{R_2}$) is a $(N-q) \times q$ matrix representing the direct (indirect) effect of TFs on non-TFs, with their elements defined previously (Eqn.\ref{eqn:Mdecomposed}-\ref{eqn:M2def}). The non-zero eigenvalues of $\textbf{M}$ are therefore the eigenvalues of the sub-matrix $\mathbf{Q}$ with elements:
\begin{equation}
    Q_{ij} = c^{ss}_i (T_{1,ij}-T_{2,ij}).
    \label{eqn:Qmatdef}
\end{equation}
When the network is sparse, each TF only regulates a small fraction of the total number of genes. Since $c^{ss} \sim 1/N$, the strength of indirect interactions are therefore typically much weaker than that of direct interactions (i.e. the non-zero elements of $\mathbf{M_2}$ are much smaller in magnitude than that of $\mathbf{M_1}$, Eqns.\ref{eqn:M1def},\ref{eqn:M2def}). 

\begin{figure}[ht]
    \centering
	\includegraphics[width=13.35cm]{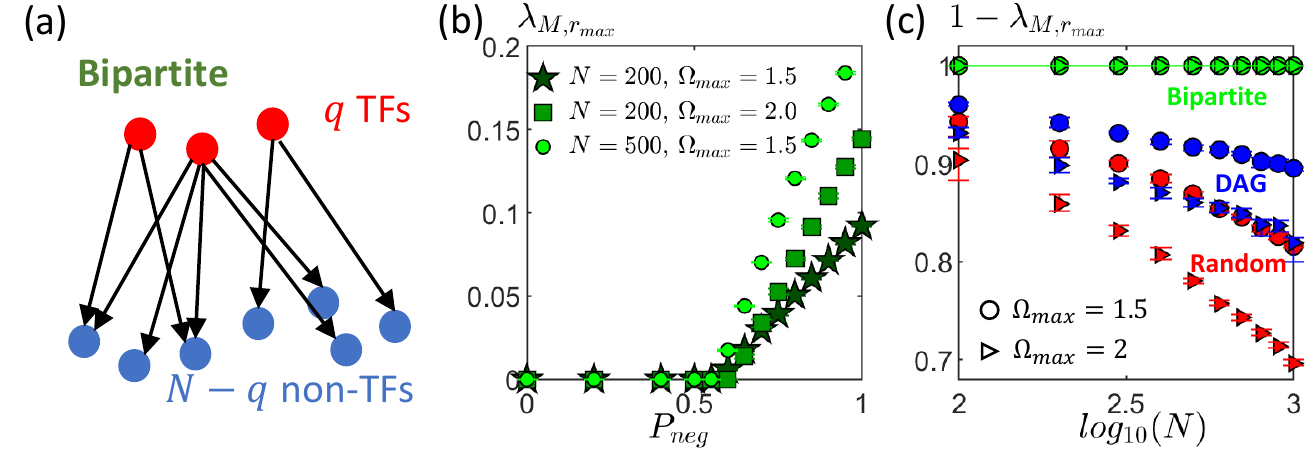}
	\caption{Stability of bipartite networks. (a) When constructing a bipartite interaction network, we group the proteins into TFs (red circles) and non-TFs (blue circles), and only allow directed regulatory interactions to go from a TF to a non-TF. (b) For bipartite networks, there is a critical value of $P_{neg}$ (that is slightly greater than $0.5$) below which $\lambda_{M,r_{max}} = 0$ and above which $\lambda_{M,r_{max}} > 0$. In the regime where $\lambda_{M,r_{max}} = 0$ (which can be considered to be `deeply' stable since it is furthest from the point $\lambda_{M,r_{max}} = 1$ where the system becomes unstable), this value of $\lambda_{M,r_{max}}$ stays the same even when $N$ (star markers to circles) or interaction strengths $\Omega_{max}$ (star markers to squares) are increased. (c) When there is an equal fraction of up/down regulatory interactions $P_{neg}=0.5$, $\lambda_{M,r_{max}}$ is independent of both $N$ and $\Omega_{max}$ for bipartite networks. This is in contrast to fully random networks (red markers) and random DAGs (blue markers) where the system approaches the instability limit ($\lambda_{M,r_{max}} = 1$) as $N$ or $\Omega_{max}$ (circles to triangles) is increased. This implies that a bipartite network structure can maintain and enhance the stability of the system as $N$ or $\Omega_{max}$ is increased. Each data point is obtained from an average of 10 randomly drawn networks, with error bars indicating the interquartile range. [Parameters: $\rho = 0.01$, $n = 1$, number of TFs $q = 0.1N$.]}
	\label{fig:BipartiteStability}
\end{figure}
% Should we include additional lines for random and DAGs in this figure?

When constructing random bipartite networks, we only allow TFs to regulate non-TFs (Fig.\ref{fig:BipartiteStability}a), implying that $\mathbf{T_1} = \mathbf{0}$. The matrix $\mathbf{Q}$ therefore only consists of weak indirect interactions, and we expect the maximal eigenvalue to be smaller than that of random networks and DAGs. Moreover, since in this case $\mathbf{Q}$ is of rank-1, it has a unique real eigenvalue $\lambda_{Q,b}$ which can be shown to be (see SI Section \ref{sec:Bipartite}):
\begin{equation}
    \lambda_{Q,b} = -\sum_{i=1}^q c_i \frac{\partial \text{log} g_T}{\partial c_i},
    \label{eqn:lambda_bipartite}
\end{equation}
where $\frac{\partial \text{log} g_T}{\partial c_i} = \sum_{j=1}^N c_j \frac{\partial \text{log} g_j}{\partial c_i}$ as defined in Eqn.\ref{eqn:M2def} are the elements of the $\mathbf{M_2}$ matrix (and therefore small when the interaction density is low). The maximum eigenvalue of the interaction matrix $\mathbf{M}$ is then given by $\lambda_{M,b} = \text{max}(\lambda_{Q,b},0)$, since $0$ is also an eigenvalue of $\mathbf{M}$ (see Eqns. \ref{eqn:M1decomposed}, \ref{eqn:M2decomposed}).

This expression (Eqn.\ref{eqn:lambda_bipartite}) implies that unlike for fully random networks and random DAGs, the stability of bipartite networks can depend strongly on the ratio of up- and down- regulating interactions (see SI Section \ref{sec:IntSignEffect}). In particular, there is a limit on the total strength of down-regulation (relative to that of up-regulation) for the system to be stable. For example, if the majority of the interactions are up-regulating, $\lambda_{Q,b}$ should be negative and hence $\lambda_{M,b}$ must be $0$. On the other hand, $\lambda_{M,b}$ must be positive when the fraction of down-regulations is sufficiently high. This tendency for inhibitory (activating) interactions to destabilize (stabilize) the system comes from the indirect effect that a regulator has on itself: a slight increase in the concentration of an inhibitor from its steady-state value will reduce the gene copy number and hence mRNA levels of the regulated gene. The mRNAs of the inhibitor therefore make up a larger fraction of the total mRNA in the cell. Since all mRNAs compete for the shared pool of ribosomes, this in turn causes the inhibitor concentrations to increase further. This positive feedback also exists in the other phases, although its physical origin may be different (see SI Section \ref{sec:IntSignEffect}, Fig. \ref{SIfig:EffectofIntSign}b). 

Indeed, by numerically constructing multiple instances of a bipartite network and varying the fraction of inhibitory interactions $P_{neg}$, we find that $\lambda_{M,b} = 0$ when $P_{neg}$ is below a critical value that is approximately (but slightly greater than) $0.5$ (Fig. \ref{fig:BipartiteStability}b). Importantly, within this regime, the value of $\lambda_{M,b}=0$ is independent of both $N$ and the strength of interactions $\Omega_{max}$ (Fig. \ref{fig:BipartiteStability}b, c). This suggests that such a bipartite network structure can help to maintain and enhance the stability of the system, especially for large $N$ and $\Omega_{max}$.

\subsubsection*{Scrambling the interactions of \textit{E. coli} transcriptional regulatory network can destabilize the system.}

Real transcription networks, however, are not strictly bipartite graphs - there are autoregulatory elements as well as transcription factors that regulate other transcription factors. To investigate how relevant network stability is to biological networks, we obtained the \textit{E. coli} transcriptional regulatory network from ref. \cite{fang2017global}. The network consists of $u = 5654$ regulatory interactions (of which $u_p = 3187$ are up-regulating), with $q = 211$ TFs regulating $N = 2274$ genes. We compared its stability with that of randomly constructed networks with the same $N$, density of interactions $\rho = \frac{u}{N^2} \approx 0.0011$, and ratio of positive (activating) to negative (inhibitory) regulation. 

We first explored two different ways of scrambling the original network: (1) randomly choosing $u$ directed connections out of the $N(N-1)$ possible connections, and (2) fixing the number of TFs $q$ and randomly choosing $u$ directed connections out of $qN$ possibilities. The second method of scrambling is motivated by the fact that $q \ll N$ and the stability of the system is governed solely by the $q \times q$ matrix $\mathbf{Q}$ representing how TFs affect TFs (Eqn.\ref{eqn:Qmatdef}). For each drawn interaction network, we randomly choose $u_p$ of the interactions to be up-regulating ($\gamma_{ij}>0$) and the rest to be down-regulating ($\gamma_{ij}<0$). We draw the fold-change $\Omega_{ij}$ of each regulatory interaction from a uniform distribution between 1 and $\Omega_{max}=1000$. This choice of $\Omega_{max}$ is motivated by the fact that TFs have been shown experimentally to change target protein levels by ~100-1000 fold \cite{kuhlman2007combinatorial}.

\begin{figure}[!ht]
    \centering
	\includegraphics[width=14.79cm]{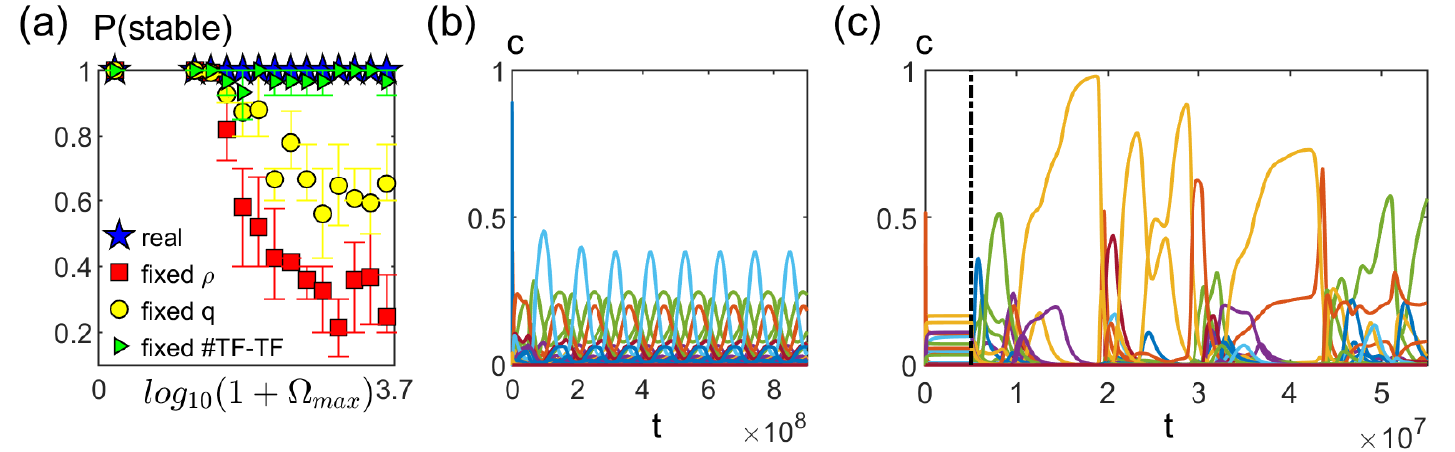}
    \caption{Comparing the \textit{E. coli} transcriptional regulatory network with random networks of the same density. (a) The actual \textit{E. coli} network does not become unstable even when the maximum regulation strength $\Omega_{max}$ is increased (blue stars). In contrast, as $\Omega_{max}$ increases, the probability of the system not having a stable fixed point increases for scrambled networks of the same interaction density $\rho = 0.0011$, regardless of whether the number of TFs $q = 211$ is kept fixed (yellow circles) or not (red squares). However, scrambling the network while maintaining the same number of TF-other TF, TF-nonTF, and self interactions can significantly enhance the probability of the system being stable (green triangles). Each of the data points are averages over 15 sets of 10 regulatory networks, with error bars indicating the interquartile range. [Other parameters: $n=2$.] (b) A typical example of oscillatory behaviour when the system no longer has a stable fixed point. [Parameters: $\Omega_{max} = 1585$, $n=2$] (c) An example of the system going unstable and exhibiting chaotic behavior when the real network is scrambled at time $t = 5 \times 10^6$ marked by the dashed vertical line. [Parameters: $\Omega_{max} = 1000$, $n=5$.] In both (b) and (c), time $t$ is in units of $1/k_p$.}
    \label{fig:realVscrambled}
\end{figure}
%In (a), the red and yellow data points are averages over 15 sets of 10 regulatory networks. The other data points are averages over 3 sets of 10 regulatory networks. 

We find that with the real network, the system always converges to a stable fixed-point regardless of the regulation strengths (Fig. \ref{fig:realVscrambled}a). In contrast, for the randomly constructed networks (both with and without keeping $q$ fixed), the probability of the system becoming unstable drastically increases when the interactions become too strong (Fig. \ref{fig:realVscrambled}a). This loss of a stable fixed point can give rise to either an oscillatory (Fig. \ref{fig:realVscrambled}b) or chaotic behaviour (Fig. \ref{fig:realVscrambled}c). This suggests that for typical regulation strengths and density, the interaction network cannot be random, and that certain structural features of real networks are important for stability. 

% \subsubsection*{A low number of TF-TF interactions helps to maintain network stability}
\subsubsection*{Network stability depends on the density of TF-TF interactions}

Since it is the maximal eigenvalue of the $q \times q$ sub-matrix $\mathbf{Q}$ (Eqn. \ref{eqn:Qmatdef}) that determines the stability of the system, and direct regulatory interactions are typically stronger than the indirect background effects, we expect a higher density of direct interactions among TFs to destabilize the system. This suggests that what matters for stability is not only the number of TFs and the total number of regulatory interactions, but also the fraction of those interactions that target TFs.   

\begin{figure}[tp]
    \centering
	\includegraphics[width=8.9cm]{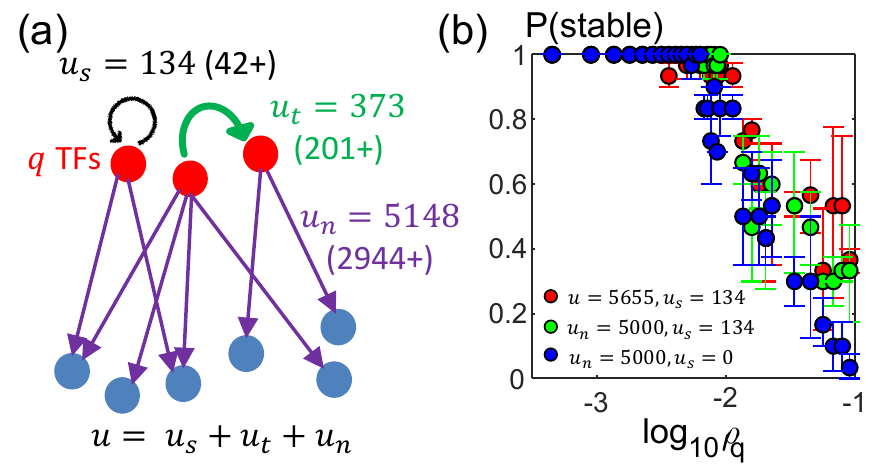}
    \caption{Effect of density $\rho_q$ of TF-otherTF interactions on stability. (a) In the real network analyzed, there are $u_s = 134$ self-regulations (of which $42$ of them are activating), $u_t = 373$ TF-other TF interactions (of which $201$ of them are activating), and $u_n = 5148$ TF-nonTF interactions (of which $2944$ of them are activating). The total number of interactions is given by $u$. (b) A randomly constructed network is almost always stable when $\rho_q$ is sufficiently low. Above a threshold value, the probability of being stable decreases with $\rho_q$. This is true with (red and green circles) or without (blue circles) self-interactions, and regardless of whether it is the total number of interactions $u$ (red circles) or the number of TF-nonTF interactions $u_n$ (green and blue circles) that is kept constant. [Parameters: $N = 2274$, $q = 211$, $n=2$, $\Omega_{max} = 1000$.]}
    \label{fig:VaryTTint}
\end{figure}

We therefore analyzed the composition of regulatory interactions in the \textit{E. coli} transcription network, and found that there are (i) $u_s = 134$ self-regulations (of which $42$ are activating), (ii) $u_t = 373$ TF-other TF interactions (of which $201$ are activating), and (iii) $u_n = u - u_s - u_t = 5148$ TF-nonTF interactions (of which $2944$ are activating) (Fig. \ref{fig:VaryTTint}a). In comparison, the scrambling method that maintained both the number of TFs and the total number of interactions gives a smaller number of self-interactions ($\langle u_s \rangle = 2.5$) and a larger number of direct TF-other TF interactions ($\langle u_t \rangle = 522$) than in the real network. 

To investigate if this could be the origin of the enhanced stability of the \textit{E. coli} regulatory network, we tried another scrambling method with the composition of the interactions kept fixed. In particular, after setting the first $q=211$ (out of $N=2274$) proteins to be TFs, we randomly drew the numbers of interaction pairs within the 3 categories (self, TF-otherTF, and TF-nonTF) by choosing each TF and its target separately. The sign of the interactions are then randomly assigned while maintaining the fraction of positive/negative interactions within each of these categories. We find that this scrambling procedure, which fixes the composition of regulatory interactions (in addition to $N$, $q$ and $\rho$), significantly increases the probability of the network having a stable fixed point (Fig. \ref{fig:realVscrambled}a).
%  (and throwing out pairs that have already been drawn)

Direct interactions among TFs can either be auto-regulatory loops or TFs regulating other TFs. We explored the effects of both of these factors, and found that assuming up- and down-regulations to be equally likely, a random network is almost always stable when the density of TF-other TF interactions $\rho_q = \frac{u_t}{q(q-1)}$ is sufficiently low (Fig.\ref{fig:VaryTTint}b). Above this threshold value of $\rho_q$, the probability of the system not exhibiting a stable steady-state increases with $\rho_q$ (Fig.\ref{fig:VaryTTint}b). This effect is observed regardless of the number of self-interactions or whether $u_n$ is kept fixed (Fig.\ref{fig:VaryTTint}b). 

While this implies that systems with a small number of TF-TF interactions are almost always stable, it does not mean that having a high density of TF-TF interactions will necessarily lead to an unstable system. This can be seen from the fact the the probability of the system being stable does not drop sharply with $\rho_q$ (Fig. \ref{fig:VaryTTint}b) $\--$ there are still systems with a relative high density of TF-TF interactions that are still stable. This suggests that in the high $\rho_q$ regime, the details of the interactions become important. For such a network with a large number of TF-TF interactions to be stable, the type and strength of those interactions will need to be more fine-tuned. 

This phenomenon that a small $\rho_q$ promotes stability is consistent with the stability of bipartite networks ($\rho_q = 0$) and the fact that direct regulatory interactions are typically much stronger than the indirect background interactions. Nevertheless, since $\mathbf{Q}$ (which has contributions from both $\mathbf{T_1}$ and $\mathbf{T_2}$, Eqn. \ref{eqn:Qmatdef}) is not a sparse matrix even when $\rho_q$ is small, we do not expect the maximal eigenvalue $\lambda_{M,r_{max}}$ to scale with $\rho_q$ the way it does for a $q \times q$ random matrix with density $\rho_q$. Indeed, we find numerically that the presence of $\mathbf{T_2}$ can affect $\lambda_{M,r_{max}}$ (Fig. \ref{SIfig:VaryTTint}), suggesting that the indirect coupling between proteins can also play a role in influencing the stability of the system. 

%(link to effect of background interactions on DAG?) 

% The fact that a small $\rho_q$ promotes stability is consistent with the stability of bipartite networks ($\rho_q = 0$). However, unlike in RMT where the maximal eigenvalue of a $q \times q$ random matrix with density $\rho_q$ scales as $\sqrt{\rho_q}$, this is not the case here (Fig. \ref{SIfig:VaryTTint}), suggesting that the coupling between TF and nonTF (captured by $\mathbf{T_2}$ in Eqn. \ref{eqn:Qmatdef}) also plays a role in influencing the stability of the system. 

%  and the significance of $\rho_q$ illustrated here may not be unexpected.

\subsection*{Effect of degradation rates on protein level stability}
So far, we have been working in the limit of fast mRNA degradation, where the stability of the system is governed only by the interaction matrix $\mathbf{M}$ (Eqn. \ref{eqn:Mmatrix}). In this regime, since $\mathbf{M}$ is independent of degradation rates $1/\tau_m$ and $1/\tau_p$ (see Eqns. \ref{eqn:Mmatrix}, \ref{eqn:phidef}, \ref{eqn:giexpression}), these do not affect whether the system is stable. The relaxation rates are also independent of $\tau_m$ and $\tau_p$, with the relaxation rate in the absence of interactions given by (from Eqn. \ref{eqn:dcpdt_smalltaum}):
\begin{equation}
    \beta_0 = k_p c_r^{ss}.
\end{equation}

However, it is not clear if this insensitivity (of both stability and relaxation rates) to $\tau_m$ and $\tau_p$ still holds outside of the $\tau_m \ll \tau_p$ regime. Within the framework of RMT, a more negative self-regulation term typically increases the relaxation rate and hence has a stabilizing effect \cite{may1972will}. Here, we ask if this is the case by investigating how mRNA and protein degradation rates affect the stability of the system and its relaxation timescale. In particular, can faster mRNA degradation rates help to stabilize a system that would otherwise be unstable if mRNAs degrade too slowly?

\subsubsection*{Values of mRNA and protein degradation rates do not affect whether the system is stable.}
To investigate how the degradation rates of proteins and mRNAs affect the stability of the system when $\tau_m$ is not too small, here we consider the full set of $2N$ equations (Eqns. \ref{eqn:dcmdt}, \ref{eqn:dcpdt}) and study how the eigenvalues of the ($2N$ $\times$ $2N$) Jacobian matrix $\bf{J}$ varies with $\tau_m$ and $\tau_p$.

To compare the relaxation rates of the full system with the protein relaxation rates when there are no interactions, we work with the transformed Jacobian matrix:
\begin{equation}
    \mathbf{\tilde{J}} =\frac{1}{\beta_0} \mathbf{J}. \label{eqn:Jtilde}
\end{equation}
% with $K = \frac{k_m k_p n_s r_s}{V} g_T$, such that in the $\tau_p = \infty$ limit, the corresponding eigenvalues $\tilde{\lambda}_\alpha$ are the actual relaxation rates relative to the growth rate of the cell. In this limit, for the cell to respond to perturbations more quickly than its growth rate, $\tilde{\lambda}_{\alpha,max}$ must be less than $-1$. But in general, the system is stable as long as $\tilde{\lambda}_{\alpha,max} < 0$. 

For an arbitrary regulatory network with a corresponding interaction matrix $\bf{M}$ (Eqn.\ref{eqn:Mmatrix}), we find that the eigenvalues $\tilde{\lambda}$ of $\mathbf{\tilde{J}}$ are given by (see SI Section \ref{sec:differentphases}):
\begin{equation}
    \tilde{\lambda} = \frac{1}{2} \left(-\omega \pm \sqrt{\omega^2 + 4\lambda_M (1+\omega)} \right) - 1, 
    \label{eqn:lambdatilde}
\end{equation}
where $\lambda_M$ are the eigenvalues of $\bf{M}$ as before, and $\omega$ is a dimensionless quantity given by:
\begin{equation}
    \omega = \frac{1}{\tau \beta_{0}},
    \label{eqn:omega}
\end{equation}
which reflects the difference between mRNA and protein degradation rates $\left(\frac{1}{\tau}=\frac{1}{\tau_m}-\frac{1}{\tau_p}\right)$. 
% in phase 1 and mRNA degradation rate in phase 3, both relative to the baseline growth rate.

\begin{figure}[tp]
    \centering
	\includegraphics[width=8.7cm]{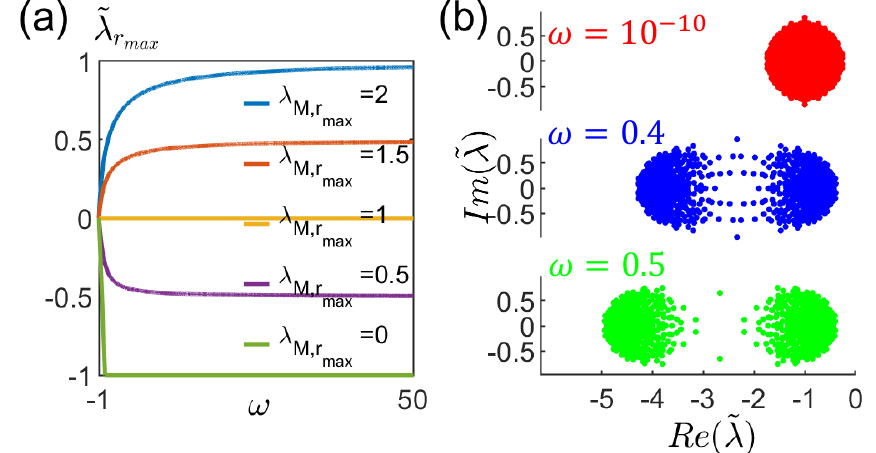}
    \caption{Effect of degradation rates on stability. (a) The system is stable if and only if $\lambda_{M,r_{max}} \leq 1$, regardless of the value of $\omega$ which increases with mRNA degradation rates (Eqn. \ref{eqn:omega}). The scaled eigenvalues $\tilde{\lambda} \to \lambda_M - 1$ in the limit of fast mRNA degradation rate $\omega \to \infty$ (Eqn. \ref{eqn:lambdatilde}). (b) Eigenvalue spectrum for different degradation rates $\tau$. When mRNA and protein degradation rates are comparable, all eigenvalues fall within a circular region (red). On the other hand, when $\tau_m \ll \tau_p$, the eigenvalue spectrum approximately resembles two circular regions, one corresponding to the dynamics of mRNAs and one for that of proteins. In this limit, increasing mRNA degradation rate only shifts the eigenvalues for the mRNA `sector' to more negative values, leaving the maximal real part of the eigenvalues approximately unchanged, $\omega = 0.5$ (green) vs $\omega = 0.4$ (blue).}
    \label{fig:EffectofTau}
\end{figure}

Since on average cell volume increases exponentially with rate (see Eqn. \ref{eqn:dpdt}):
\begin{equation}
    \mu = k_p \phi_r - \frac{1}{\tau_p},
    \label{eqn:mu1}
\end{equation}
a growing cell has to satisfy the condition $\frac{1}{\tau_p k_p \phi_r} < 1$. Therefore, since $\tau_m \geq 0$, we have $\omega \geq -1$. The expression for $\tilde{\lambda}$ (Eqn. \ref{eqn:lambdatilde}) therefore implies that the system is stable if and only if $\lambda_{M,r_{max}} \leq 1$, regardless of the value of $\tau_m$ and $\tau_p$ (Fig \ref{fig:EffectofTau}a). We find that despite differences in the details of the model, this conclusion still holds in the other phases (see SI Section \ref{sec:differentphases}).

Therefore, unlike what has been argued in the literature and what one might expect from RMT, changing mRNA nor protein degradation rates has no effect on whether the overall system is stable. If steady-state protein concentrations are unstable because $\lambda_{M,r_{max}}$ is too large (e.g. when interactions are too strong), increasing mRNA or protein degradation rates can never help to stabilize the system. 

Importantly, this finding also implies that our results for how structural features of the transcription network affects stability holds outside the regime of fast mRNA degradation, since stability only depends on $\mathbf{M}$.

% In phase 1 of the model, a growing cell has to satisfy the condition $\frac{1}{\tau_p k_p \phi_r} < 1$ (Eqn.\ref{eqn:mu1}). Therefore, since $\tau_m \geq 0$, $\omega_\alpha \geq -1$ in both phases. With the expression for 

\subsubsection*{Increasing mRNA degradation rate can improve response times, but only up to some limit.}

Besides system stability, another quantity of biological interest is the response time of the system to perturbations, which is especially relevant for cells experiencing changes in nutrient conditions \cite{van2014lost, erickson2017global}. Since this relaxation timescale is determined by the slowest eigenvalue of the Jacobian matrix, here we discuss how the maximal real part of the eigenvalues $\tilde{\lambda}_{r_{max}}$ changes with $\tau$.

The expression for $\tilde{\lambda}$ (Eqn.\ref{eqn:lambdatilde}) implies that when the system is stable ($\lambda_{M,r_{max}} < 1$), the rate at which the system relaxes to steady-state initially increases as $\omega$ increases from $-1$, but eventually plateau off $\--$ in the $\omega \to \infty$ limit (where $\tau_m \ll \tau_p$), $\tilde{\lambda} \to \lambda_M - 1$ (Eqn. \ref{eqn:lambdatilde}, Fig. \ref{fig:EffectofTau}a). This implies that there is some benefit to having fast mRNA degradation in terms of response times, but once mRNA degrades much faster than proteins, further increasing mRNA degradation rate no longer affects the response time of the system. The eigenvalue spectrum in this $\tau_m \ll \tau_p$ limit appears to consist of two circular regions, one for the dynamics of mRNAs and the other for that of proteins (Fig. \ref{fig:EffectofTau}b), reminiscent of the RMT's circular law. Increasing $\tau_m$ only shifts the eigenvalues corresponding to the mRNA sector and hence does not affect $\tilde{\lambda}_{r_{max}}$. This is consistent with the fact that when $\tau_m \ll \tau_p$, the dynamics of the overall system is governed only by the protein `sector' (Eqn.\ref{eqn:dcpdt_smalltaum}). Therefore, the slowest relaxation rate back to steady-state levels depends only on $\mathbf{M}$ and increasing mRNA degradation rate no longer improves the response time.

\section*{Discussion} 
In systems with a large number of interacting components, the question of stability is often an important one, as results from random matrix theory (RMT) predict instability when the system size $N$ becomes too large or interactions become too strong. In the context of gene expression, transcriptional regulation is crucial for cells to adapt to different environmental conditions by changing their gene expression levels. It is therefore important for transcriptional regulatory networks (TRNs) to be able to accommodate a large number of regulatory interactions without the system going unstable. However, we find here that similar to the intuition provided by RMT, $\lambda \sim \sqrt{N}$ for a fully random regulation network, suggesting that the system will go unstable as the number of genes exceeds a threshold. In fact, based on typical values for the density of actual regulatory networks and interaction strengths, we find that the system has a high probability of being unstable if the TRN is randomly constructed.

Besides the number of genes, and the density and strengths of interactions, there are other factors that can affect the stability of the system, one of which is the network topology. This aspect is particularly relevant in this system since TRNs are far from being random but instead consist of recurring motifs. While the properties of these specific motifs have been widely studied and shown to be important for specific functions such as adaptation, robustness, and fast response to environmental changes \cite{alon2019introduction, shen2002network, alon2007network}, how they contribute to the overall stability of the network is less clear. We find here that global structural features of the network, which are fundamentally shaped by many of these motifs, can play a huge role in determining the stability of the system. In particular, given the same number of proteins, TFs, interaction density and regulation strengths, a network that resembles a bipartite graph with a lower density of TF-otherTF interactions $\rho_q$ has a higher chance of being stable. The significance of $\rho_q$ fundamentally arises because of two main factors: (i) the eigenvalues of the Jacobian matrix and hence the stability of the system about its steady-state are governed only by the TF sector (i.e. how perturbations in TF concentrations affect TFs), and (ii) for a sparse regulatory network, the indirect background interactions arising from competition for ribosomes between different genes are typically much weaker than the direct regulatory interactions.
% Such a network would resemble a bipartite graph, in which all interactions involve a TF regulating a nonTF. 

TRNs are also known to be scale-free, having a power-law out-degree distribution. This is consistent with the fact that most TFs only regulate a small number of genes, but there are TFs that regulate a very large number of genes (`master regulators'). Within a more abstract model of gene regulatory dynamics, the presence of these outgoing hubs has been shown to significantly increase the probability of the system reaching a stable target phenotype when the interaction strengths are allowed to vary while the network topology is kept fixed \cite{schreier2017exploratory}. Here, we find that having a low $\rho_q$ can already significantly stabilize the system without the need to control the degree distributions. Nevertheless, having just a few master regulators may contribute to the network having a low $\rho_q$ if for instance most of the regulations on TFs are carried out by the master regulators (and non-master regulators predominantly regulate non-TFs). % Alternatively, power law out-degree distribution may be useful for other functions.. (i'm sure there are lots of theories out there for why scale free networks are good).

Besides structural features of the network, another factor that could affect stability is the degradation rates of mRNA and proteins. Based on RMT, one may expect faster degradation to stabilize the system. This has in fact been argued to be the case \cite{chen2019gene, zhao2017weak}. However, by taking into account the dynamics of protein concentrations and how it couples to the dynamics of mRNA levels, we find that this is not the case. Instead, the stability of the system depends solely on the regulatory network and the strengths of those regulations $\--$ if the system is unstable, it will be unstable regardless of how fast mRNA or protein degrades. This highlights the importance of taking into account key aspects of the interactions (through the form of the dynamical equations) when analyzing the stability of large coupled systems, similar in spirit to studies of ecological models where explicitly considering interactions mediated through competition for nutrients can give drastically different results from assuming random pairwise interactions between species \cite{cui2019diverse}. This prediction can also potentially be tested in the lab by varying the degradation rates of mRNAs (e.g. by using genetically modified RNases) or proteins (e.g. by using genetically modified proteases) in the cell and observing the dynamics of protein concentrations.

% Relate back to ecology example again?
% say something about how it's important to take into account key qualitative aspects of the model, but our results are robust to the 3 phases although the details of the different phases are quite different?

From an evolutionary perspective, there are many possible factors (such as the range of gene expression levels, environmental conditions, response time \cite{van2014lost, erickson2017global}, level of unwanted crosstalk \cite{friedlander2016intrinsic}, etc.) that drive the addition or removal of regulatory connections. Our findings suggest that in addition to these considerations, another fundamental factor is the stability of the overall network. For example, there could be many ways of achieving a certain task such as allowing the cell to switch between two desired gene expression levels in two different nutrient conditions, but the only ones that can survive are those that also maintain the stability of the system. In other words, stability of the system may have played a role in shaping current existing regulatory networks through the evolutionary process. Our approach can therefore provide insights into the design and evolutionary constraints for a functional regulatory network, which may potentially be useful for guiding the construction of synthetic genetic circuits \cite{adamala2017engineering, ellis2009diversity, noireaux2011development}. In the future, the ability to experimentally engineer a large, random regulatory circuit within cells could also allow testing of the results we have described.  
% For example, our results suggest that if one were to experimentally construct an artificial cell containing a large number of genes with strong, random regulatory interactions between them, there is a high chance of observing oscillatory or chaotic dynamics in the concentrations of all proteins.

% The dynamics of large, multi-component networks within cells has previously been studied in the context of protein-protein interaction network. 
 In addition to transcriptional regulation, gene expression is also regulated at the post-transcriptional (e.g. through small-RNAs or micro-RNAs) and post-translational (e.g. through post-translational modifications) level. Our framework can be extended to take into account these effects (see SI Section \ref{sec:PTM} for an example). How stability of the system is affected by the coupling between these different forms of regulation with potentially different network structures is an interesting question that we leave for future work. Besides stability (determined by the eigenvalues of $\mathbf{J}$), in the future it could also be instructive to investigate the spread of perturbations within the regulatory network (i.e. the eigenvectors of $\mathbf{J}$). This is analogous to the study of how concentration perturbations propagate in protein-protein interaction networks within the cell \cite{maslov2007spreading}.

% To illustrate the importance of structural features to the stability of the system, we made use of \textit{Ecoli.} TRN. While this is probably the best characterized TRN, it remains incomplete. 

% - comment on the how negative TF-TF interactions help stability but negative TF-nonTF interactions destabilizes the system.

%While there may not be an specific motif that enhances the stability of the system, our findings provide insights into design and evolutionary constraints for a functional regulatory network.

\section*{Data Availability}
The data that support the findings of this study are available from the corresponding
author upon request.

\section*{Code Availability}
All codes can be found on GitHub repository https://github.com/yipeiguo/TRNstability.

\bibliographystyle{ieeetr}
\bibliography{MainTextRef}

\section*{Acknowledgments}
We thank Rui Fang, Jie Lin, Haim Sompolinsky, Grace Zhang, David Nelson, Naama Brenner and Guy Bunin for useful discussions and feedback. This research was supported by the National Science Foundation through MRSEC DMR 14-20570, the Kavli Foundation, and the NSF CAREER 1752024. 

\section*{Author Contributions}
Y.G., A.A. designed research, performed research, and wrote the paper. 

\section*{Competing Interests}
All authors declare that they have no competing interests. 

\newpage 

\renewcommand{\thefigure}{S\arabic{figure}}
\renewcommand{\theequation}{S\arabic{equation}}
\renewcommand{\thetable}{S\arabic{table}}
\setcounter{figure}{0}  
\setcounter{equation}{0}  

\section*{Supplementary Information}
\appendix

\section{Example of a thermodynamic model of RNA polymerases and transcription factors binding to DNA}
\label{sec:ThermoModel}

We consider the scenario where a gene has 1 promoter site and $L$ regulatory sites, each corresponding to a binding site for a different transcription factor.

Let $q_i = \frac{c_i}{K_i} = e^{-\beta(\epsilon_i - \mu_i)}$ be the binding affinities of each site $i=0,1,...,L$, where $c_i$ is the concentration of the protein $i$, $K_i$ and $\epsilon_i$ are respectively the dissociation constant and binding energy between protein $i$ and site $i$, and $\mu_i$ is the chemical potential of $i$. We choose the index $i=0$ to represent binding of RNA polymerase to the promoter and the other indices represent TF binding to the corresponding regulatory site. 

The state of the system is then given by $\vec \sigma$ with $\sigma_i = \{0,1\}$ representing whether the $i^{th}$ binding site is occupied. We allow pairwise interactions between RNAP and each of the TFs, but neglect any pairwise interactions among the TFs, such that the free energy $E$ of any state is given by:
\begin{equation}
    E(\vec \sigma) = \sum_i (\epsilon_i-\mu_i) \sigma_i - \sum_j \frac{\text{log} w_{0j}}{\beta} \sigma_0 \sigma_j,
\end{equation}
where $w_{0j} \geq 0$ captures the strength and nature of the pairwise interaction between a bound RNAP and a bound TF $j$. Specifically, $w > 1$ indicates a positive interaction (with the TF up-regulating gene expression), $w = 1$ indicates no interaction, while $w < 1$ indicates a repulsive interaction. The limit where $w_{0j}=0$ corresponds to the case where the TF is a steric inhibitor i.e. binding of $j$ blocks RNAP from binding to the promoter. 

Denoting $Z^{ON} (Z^{OFF})$ as the sum over the weights of all possible RNAP-bound `ON' (RNAP-unbound `OFF') configurations, the equilibrium probability $P_b$ of RNAP binding to the promoter is given by
\begin{equation}
\begin{split}
    P_b &= \frac{Z^{ON}}{Z^{ON}+Z^{OFF}} \\
    &= \frac{q_0 \prod_{i=1}^L (1+w_{0i}q_i)}{\prod_{i=1}^L (1+q_i) + q_0 \prod_{i=1}^L (1+w_{0i}q_i)} \\
    &= P_{b0} F_{reg}(\vec{c}) ,
\end{split}
\end{equation}
where $P_{b0} = \frac{q_0}{1+q_0}$ is the probability of RNA polymerase being bound to the promoter in the absence of any transcriptional regulation ($L=0$), and $F_{reg}(\vec{c})$, which is the regulatory function which captures the effect of TFs on the the binding of RNAP to the promoter, is given by:
\begin{equation}
\begin{split}
    F_{reg}(\vec{c}) &= \frac{(1+q_0) \prod_{i=1}^L \left(\frac{1+w_{0i}q_i}{1+q_i} \right)}{1+q_0 \prod_{i=1}^L \left(\frac{1+w_{0i}q_i}{1+q_i} \right)} \\
    &\approx \prod_{i=1}^L \left(1+\frac{(w_{0i}-1)q_i}{1+q_i} \right),
\end{split}
\end{equation}
with the approximation taken in the limit of low RNAP concentrations $q_0 \ll \prod_{i=1}^L \left(\frac{1+q_i}{1+w_{0i}q_i} \right)$ and $q_0 \ll 1$.

This model is therefore an example of how a multiplicative form for $F_{reg}(\vec{c})$ can arise, and serves as a motivation for our choice of regulatory function for the effective gene copy number (which we assume to be proportional to the probability of RNAP binding to promoter). Even though in this model the Hill coefficient is $1$ for the effect of individual TFs, one could imagine higher Hill coefficients if there are cooperative effects in the binding of each TF to its binding site.

\section{Summary of model parameters}
\label{sec:ParameterList}

\begin{table}[h]
\begin{center}
 \begin{tabular}{|c | c | c | c |}
 \hline
 Parameter & Definition/Description & How value is set in simulations \\ [0.5ex] 
 \hline\hline
 \multicolumn{3}{|l|}{\textit{Transcription and transcriptional regulation}} \\
 \hline
 $k_m$ & transcription rate of a single RNA polymerase & N/A (fast mRNA degradation limit)  \\ 
 \hline
 $\tau_m$ & mRNA lifetime & N/A (fast mRNA degradation limit)  \\
 \hline
 $g_i$ & effective gene copy number of $i$ & $g_i(\vec{c}) = g_{i0} \prod_j (1 + \gamma_{ij} f_{ij}(c_j))$ (Eqn.~\ref{eqn:giexpression})  \\
 \hline
 $g_{i0}$ & effective gene copy number of $i$ if  & drawn from a uniform distribution  \\
 & it were unregulated & between 0 and 1 \\
 \hline
 $\phi_i$ & gene allocation fraction of $i$ & \rule{0pt}{15pt} $\frac{g_i}{\sum_k g_k}$ (Eqn.~\ref{eqn:phidef}) \\ [2ex]
 \hline
 $\phi_{i0}$ & gene allocation fraction of $i$ & \rule{0pt}{15pt} $\frac{g_{i0}}{\sum_k g_{k0}}$  \\
 & without any regulatory interactions & \\
 \hline
 $\gamma_{ij}$ & controls the type and strength of regulation & $\gamma_{ij} = \begin{cases}
    \Omega_{ij} - 1 & \text{if $\gamma_{ij} > 0$}
    \\
    \frac{1}{\Omega_{ij}} - 1 & \text{if $\gamma_{ij} < 0$}
    \end{cases}$ (Eqn.~\ref{eqn:gammadef})\\ 
 & & \\
 \hline
 $\Omega_{ij}$ & fold-change of each regulatory interaction, & drawn from a distribution $P(\Omega)$   \\ 
 & controls strength of interaction & (either uniform or log-uniform)  \\
 & & between 1 and $\Omega_{max}$ \\
 \hline
 $f_{ij}$ & How protein $j$ affects gene copy number of $i$ & \rule{0pt}{20pt} $f_{ij}(c_j) =  \frac{c_j^{n_{ij}}}{K_{ij}^{n_{ij}}+c_j^{n_{ij}}}$ (Eqn.~\ref{eqn:fijexpression}) \\ [3ex]
 % & & \\
 \hline
 $n_{ij}$ & Hill coefficient of $f_{ij}$ & same constant ($n = 1, 2$ or $5$) for all $i,j$ \\
 \hline
 $K_{ij}$ & concentration of $j$ at which $f_{ij} = 0.5$ & set to be $\phi_{j0}$  \\ 
 \hline\hline
 \multicolumn{3}{|l|}{\textit{Translation}} \\
 \hline
   &  & \multirow{3}{16em}{constant, time is measured in units of $1/k_p$.}  \\ 
  $k_p$ & translation rate of a single ribosome & \\
  & & \\
 \hline
  &  & \multirow{3}{16em}{N/A, does not affect dynamics of protein concentrations in the limit of fast mRNA degradation.}    \\
 $\tau_p$ & protein lifetime & \\
 & & \\
 \hline
\end{tabular}
\caption{List of model parameters, their definitions and how they are chosen in the simulations.\label{tab:params}}
\end{center}
\end{table}

\section{Effect of stochasticity in gene expression and during cell division}
\label{sec:Gillespie}

\begin{figure}[!htb]
    \centering
	\includegraphics[width=8.7cm]{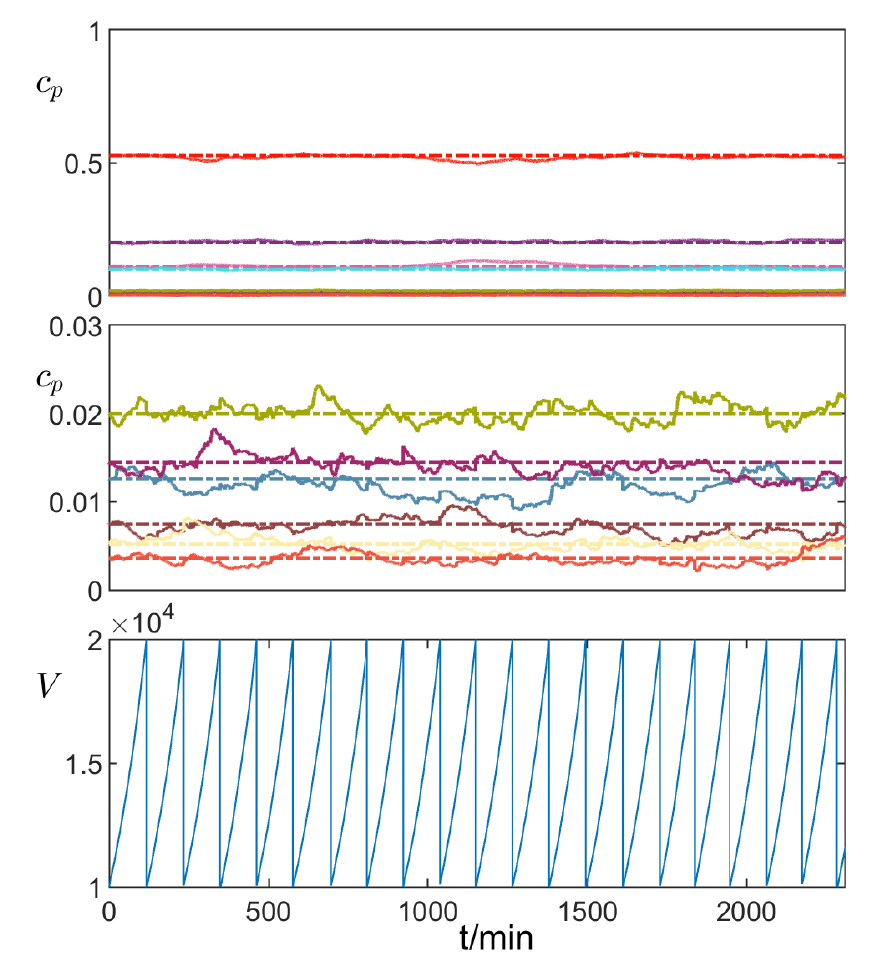}
    \caption{Gillespie simulations of gene expression and binomial sampling of mRNA and protein numbers during cell division. The protein concentrations obtained from a single simulation (solid lines in top, middle panels) fluctuate around the steady-state solution obtained from deterministic dynamical equations (dashed horizontal lines, obtained from solving main text Eqns.\ref{eqn:dcmdt}, \ref{eqn:dcpdt}).  Different colors represent different proteins, the middle panel is a zoomed-in view of the top panel in the low $c_p$ regime. The bottom panel shows the corresponding changes in cell volume, with vertical lines indicating cell division events, at which point the number of mRNA and protein molecules in the daughter cell is sampled from a binomial distribution.  [Parameters: $N = 10$, random interaction network with $\rho = 0.5$, $\Omega_{max} = 20$, $n = 1$, $k_m = 0.3$, $k_p = 0.03$, $\tau_m = 3.5$, $\tau_p = 1 \times 10^8$] }
    \label{SIfig:Gillespie}
\end{figure}

To explore the effect of stochasticity in gene expression and binomial sampling of molecules during cell division, we carry out Gillespie simulations. In these simulations, we keep track of the number of mRNA and protein molecules of each gene $i = 1,..., N$. At every time step, the production rate of each mRNA is given by $\Gamma_{mi} = k_m \phi_i(\vec{c})$, the production rate of each protein is given by $\Gamma_{pi} = k_p \frac{p_i}{\sum_k p_k} r$, and the degradation rates of each mRNA and protein are given by $\frac{m_i}{\tau_m}$ and $\frac{p_i}{\tau_p}$ respectively. The next event and the time to the next event are then drawn based on these rates. We assume that the cell divides whenever the volume (i.e. total protein number) reaches a threshold value ($2 \times 10^4$ in Fig.~\ref{SIfig:Gillespie}). When the protein number is large and the system is stable, we find that protein concentrations fluctuate around the steady-state solution obtained from the corresponding deterministic dynamical equations (Fig. \ref{SIfig:Gillespie}).

\section{Dynamics and stability of protein concentrations in different phases}
\label{sec:differentphases}

\subsection{Phase 1: The regime where both RNAPs and ribosomes are limiting ($n < n_c$, $r < r_c$)}

When both RNAPs and ribosomes are limiting, the dynamics of mRNA $c_{mi}$ and protein $c_i$ concentrations within the cell are governed by the following equations (Eqns. \ref{eqn:dcmdt} and \ref{eqn:dcpdt} of the main text):
\begin{equation}
    \frac{dc_{mi}}{dt} = k_m \phi_i({c}) c_n - c_{mi} \left(k_p c_r + \frac{1}{\tau}\right)
\end{equation}
\begin{equation}
    \frac{dc_i}{dt} = k_p c_r \left( \frac{c_{mi}}{c_{mT}} - c_i \right),
\end{equation}
where $k_m$ and $k_p$ are constants characterizing the transcription and translation rates of a single RNA polymerase and ribosome respectively, $\phi_i(\vec{c})= \frac{g_i(\vec{c})}{\sum_j g_j(\vec{c})}$ is the gene allocation fraction (with $g_i(\vec{c})$ being the effective copy number of gene $i$), $c_{mT} = \sum_i c_{mi}$ and $\frac{1}{\tau} = \frac{1}{\tau_m} - \frac{1}{\tau_p}$, with $\tau_m$ and $\tau_p$ being the lifetimes of mRNA and proteins respectively. 

The corresponding steady-state concentrations are given by:
\begin{equation}
    c_{mi,ss} = \frac{k_m c_n^{ss}}{k_p c_r^{ss} + \frac{1}{\tau}} \phi_i(\vec{c}^{ss})
    \label{eqn:cm_ss}
\end{equation}
\begin{equation}
    c_i^{ss} = \phi_i(\vec{c}^{ss}). 
    \label{eqn:c_ss_phase1}
\end{equation}
Since by definition $\sum_i \phi_i = 1$, the total steady-state mRNA concentration is
\begin{equation}
    c_{mT,ss} = \frac{k_m \phi_n}{k_p \phi_r + \frac{1}{\tau}}. 
    \label{eqn:cmT}
\end{equation}

We denote the total number of genes by $N$, and choose the index $N-1$ to represent polymerases (the number of which we also denote as $n$) and the $N^{th}$ index to represent ribosomes (the number of which we also denote by $r$). The Jacobian of the full coupled mRNA-protein system is a $2N \times 2N$ matrix $\bf{J} = \left[\begin{array}{cc}
\bf{A} & \bf{B} \\
\bf{C} & \bf{D} \\
\end{array} \right]$, where $\textbf{A} = -\left(\frac{1}{\tau}+k_pc_r\right) \bf{I}$ is the $N \times N$ matrix representing how mRNA concentrations affect one another, and $\textbf{D} = -k_pc_r \bf{I}$ is the $N \times N$ matrix representing how protein concentrations affect one another. Since $c_i^*$'s are independent of $\tau$ (Eqn. \ref{eqn:c_ss_phase1}), it is convenient to define
\begin{equation}
    \mathbf{\tilde{J}} = \frac{1}{k_p c_r} \mathbf{J} 
    \label{eqnSI:Jtilde}
\end{equation}
such that the system is stable if and only if the maximal real part of the eigenvalues of $\bf{\tilde{J}}$ is less than 0. 
The elements of $\bf{\tilde{J}}$ are then given by $\bf{\tilde{J}} = \left[\begin{array}{cc}
\bf{\tilde{A}} & \bf{\tilde{B}} \\
\tilde{\bf{C}} & \bf{\tilde{D}} \\
\end{array} \right]$, with $\tilde{\textbf{A}} = -\left( \frac{1}{\tau k_p c_r}+1 \right) \bf{I}$, $\tilde{\bf{D}} = -\bf{I}$, and
\begin{equation}
    \tilde{B}_{ij} = \begin{cases}
    \frac{k_m \phi_n}{k_p \phi_r} \frac{\partial \phi_i}{\partial c_j}, & \text{for $j=1,2,...,N-2$} \\
    \frac{k_m \phi_i}{k_p \phi_r}, & \text{for $j=N-1$} \\
    -\frac{\phi_i}{\phi_r} c_{mT}, & \text{for $j=N$}
    \end{cases}
\end{equation}
\begin{equation}
    \tilde{C}_{ij} = \begin{cases}
    \frac{1-\phi_i}{c_{mT}}, & \text{for $j=i$} \\
    -\frac{\phi_i}{c_{mT}}, & \text{for $j \neq i$}
    \end{cases}
\end{equation}
where we have made the assumption that RNAPs and ribosomes cannot act as transcription factors.

Let $\tilde{\lambda}$ be the eigenvalues of $\bf{\tilde{J}}$ with corresponding eigenvectors  $\overrightarrow{v} = \left(\begin{array}{c}
\overrightarrow{v_1}  \\
\overrightarrow{v_2}  \\
\end{array} \right)$. Then $\tilde{\bf{A}} \overrightarrow{v_1} + \tilde{\bf{B}} \overrightarrow{v_2} = \tilde{\lambda} \overrightarrow{v_1}$ and $\tilde{\bf{C}}\overrightarrow{v_1} = (\tilde{\lambda}+1) \overrightarrow{v_2}$, which gives $\tilde{\textbf{C}}\tilde{\textbf{B}} \overrightarrow{v_2} = (\tilde{\lambda}+1) \left(\tilde{\lambda} + \frac{1}{\tau k_p c_r} + 1 \right) \overrightarrow{v_2}$, where the elements of $\bf{\tilde{C}\tilde{B}}$ are given by:
\begin{equation}
\begin{split}
    (\tilde{C}\tilde{B})_{ij} &= \sum_k \tilde{C}_{ik} \tilde{B}_{kj} \\ 
        &= \begin{cases}
    \frac{k_m \phi_n}{k_p \phi_r c_{mT}} \left(\frac{\partial \phi_i}{\partial c_j} - \phi_i \sum_k \frac{\partial \phi_k}{\partial c_j} \right), & \text{for $j=1,2,...,N-2$} \\
    \frac{k_m}{k_p c_{mT}} \left( \frac{\phi_i}{\phi_r} - \phi_i \sum_k \frac{\phi_k}{\phi_r} \right), & \text{for $j=N-1$} \\
    -\frac{\phi_i}{\phi_r} + \phi_i \sum_k \frac{\phi_k}{\phi_r}, & \text{for $j=N$}
    \end{cases} \\
    &= \begin{cases}
    \left(1+\frac{1}{\tau k_p \phi_r} \right) \frac{\partial \phi_i}{\partial c_j}, & \text{for $j=1,2,...,N-2$} \\
    0, & \text{for $j=N-1,N$} \\
    \end{cases}
\end{split}
\end{equation}

This therefore provides a relation between each value of $\tilde{\lambda}$ and its corresponding eigenvalue of $\frac{\partial \phi}{\partial c}$, which we denote by $\lambda_M$. Since $\lambda_M$ is independent of $\tau$, we can find how $\tau$ affects $\tilde{\lambda}$ for any given $\lambda_M$:
\begin{equation}
    \tilde{\lambda} = \frac{1}{2}\left(-\omega \pm \sqrt{\omega^2 + 4 \lambda_M (1+\omega)} \right) - 1,
    \label{eqn:lambda_phase1}
\end{equation}
where $\omega = \frac{1}{\tau k_p c^*_r}$.

\subsection{Phase 2: The regime where RNAPs are in excess and ribosomes are limiting ($n \geq n_c$, $r < r_c$)}

Whenever it is the gene copy numbers (instead of RNAPs) that are limiting ($n \geq n_c$), the transcription rate is no longer proportional to the number of RNAPs, and hence it is the mRNA numbers $m_i$ rather than their concentrations that are kept at steady-state levels within the cell. We therefore analyze the the dynamics for $m_i$ and $c_i$ which in this case are given by:
\begin{equation}
    \frac{dm_i}{dt} = k_m g_i({c})n_s - \frac{m_i}{\tau_m}
\end{equation}
\begin{equation}
    \frac{dc_i}{dt} = k_p c_r \left( \frac{m_i}{m_T} - c_i \right),
\end{equation}
where $m_T = \sum_j m_j$ is the total number of mRNAs, $n_s$ is the maximal number of RNA polymerases a single gene can accommodate, and the other variables are as defined previously.

The corresponding steady-state mRNA and protein levels are:
\begin{equation}
    m_i^{ss} = k_m n_s g_i(\vec{c}^{ss}) \tau_m 
\end{equation}
\begin{equation}
    c_i^{ss} = \frac{m_i^{ss}}{m_T^{ss}} = \phi_i(\vec{c}^{ss}),
\end{equation}
where we note that as before the steady-state protein concentrations are independent of the degradation lifetimes.

Following the same approach as in the previous section, we define the scaled Jacobian matrix (Eqn. \ref{eqnSI:Jtilde}), where now the elements of $\bf{\tilde{J}}$ are given by $\tilde{\textbf{A}} = \left(-\frac{1}{\tau_m k_p c_r}\right) \bf{I}$, $\tilde{\bf{D}} = -\mathbf{I}$
\begin{equation}
    \tilde{B}_{ij} = \frac{k_m n_s}{k_p c_r} \frac{\partial g_i}{\partial c_j}
\end{equation}
\begin{equation}
    \tilde{C}_{ij} = \frac{1}{m_T} (\delta_{ij} - c_i^*),
\end{equation}
such that
\begin{equation}
\begin{split}
    (\tilde{C}\tilde{B})_{ij} &= \frac{1}{\tau_m k_p c_r} \frac{1}{g_T} \left(\frac{\partial g_i}{\partial c_j} - c_i \sum_k \frac{\partial g_k}{\partial c_j} \right) \\
    &= \frac{1}{\tau_m k_p c_r} \frac{\partial \phi_i}{\partial c_j}.
\end{split}
\end{equation}
The eigenvalues $\tilde{\lambda}$ of $\bf{\tilde{J}}$ are hence given by 
\begin{equation}
    \tilde{\lambda} = \frac{1}{2}\left(-\omega_2 \pm \sqrt{\omega_2^2 + 4 \lambda_M (\omega_2+1)} \right) - 1,
    \label{eqn:lambda_phase2}
\end{equation}
where $\omega_2 = \frac{1}{\tau_m k_p c_r^{ss}} - 1$, and as before $\lambda_M$ denote the eigenvalues of $\frac{\partial \phi}{\partial c}$, which are independent of $\tau$. This equation is the same as that in phase 1 (Eqn. \ref{eqn:lambda_phase1}) with $\omega$ replaced by $\omega_2$. 

Therefore, in both of these phases we get similar dependence of the stability of the system on degradation rates - the system is always stable as long as $\lambda_M < 1$ and unstable if $\lambda_M > 1$, regardless of the values of $\omega$ or $\omega_2$.

\subsection*{Phase 3: The regime where both RNAPs and ribosomes are in excess ($n \geq n_c$, $r \geq r_c$)}

In this regime, the dynamics of $m_i$ and $c_i$ are given by:
\begin{equation}
    \frac{dm_i}{dt} = \tilde{k_m} g_i(\vec{c}) - \frac{m_i}{\tau_m}
\end{equation}
\begin{equation}
    \frac{dc_i}{dt} = \tilde{k_p} (m_i - c_i m_T), 
\end{equation}
where $\tilde{k_m} = k_m n_s$, and $\tilde{k_p} = \frac{k_p r_s}{V}$ is dependent on cell volume which is linearly increasing over time. It is useful to define the growth rate per unit volume $\mu_3$, which is given by:
\begin{equation}
    \mu_3 = \frac{k_m k_p r_s n_s \tau_m g_T}{V} - \frac{1}{\tau_p}.
    \label{eqn:mu3}
\end{equation}

At steady-state,
\begin{equation}
    m_i^{ss} = \tilde{k_m} g_i(\vec{c}^{ss}) \tau_m 
\end{equation}
\begin{equation}
    c_i^{ss} = \frac{m_i^{ss}}{m_T^{ss}} = \phi_i(\vec{c}^{ss}),
\end{equation}
and while these are constant over the whole cell cycle, the rate at which the system goes back to steady-state levels after a perturbation depends on its current volume at that point in time. 

The Jacobian matrix of this system can again be written as $\bf{J} = \left[\begin{array}{cc}
\bf{A} & \bf{B} \\
\bf{C} & \bf{D} \\
\end{array} \right]$, where $\textbf{A} = -\frac{1}{\tau_m} \bf{I}$, $\textbf{D} = -\tilde{k}_1 m_T \bf{I}$, $B_{ij} = \tilde{k_m} \frac{\partial g_i}{\partial c_j}$, and $C_{ij} = \tilde{k_p}(\delta_{ij}-c_i)$. Unlike phases 1 and 2, here $\textbf{D}$ depends on $m_T = \tilde{k_m} g_T \tau_m$ which is a function of $\tau_m$. 

The eigenvalues $\lambda$ of $\bf{J}$ can be found from
\begin{equation}
    \tilde{\textbf{C}}\tilde{\textbf{B}} \overrightarrow{v_2} = \left(\lambda+\frac{1}{\tau_m} \right) \left(\lambda + \tilde{k}_1 m_T \right) \overrightarrow{v_2},      
\end{equation}
with
\begin{equation}
\begin{split}
    (CB)_{ij} &= \tilde{k}_0 \tilde{k}_1 \left(\frac{\partial g_i}{\partial c_j} - c_i \sum_k \frac{\partial g_k}{\partial c_j} \right) \\
    &= K \frac{\partial \phi_i}{\partial c_j},
\end{split}
\end{equation}
where $K = \tilde{k}_0 \tilde{k}_1 g_T$. Unlike the other phases, here we choose not to scale $\bf{J}$ by the diagonal elements of $\textbf{D}$ since we are investigating the effect of $\tau_m$ on the eigenvalues and $\bf{D}$ itself depends on $\tau_m$.

We therefore have
\begin{equation}
    \lambda = \frac{-\left(\frac{1}{\tau_m}+K\tau_m\right) \pm \sqrt{\left(\frac{1}{\tau_m}+K\tau_m\right)^2 - 4K(1-\lambda_M)}}{2},
\end{equation}
where as before, $\lambda_M$ are the eigenvalues of the interaction matrix $\mathbf{M} = \frac{\partial \phi}{\partial c}$.

This implies that the system becomes marginally stable ($\lambda \to 0$), for both $\tau_m \rightarrow 0$ and $\tau_m \rightarrow \infty$, i.e. in both these limits, even if the system is stable, it takes a long time for it to relax back to its steady-state when perturbed. This suggests that there is an intermediate regime of $\tau_m$ for which the system responds fast to perturbations away from steady-state. This `Goldilocks effect' arises because when $\tau_m$ is large, the restoring force for mRNA numbers is small, while for small $\tau_m$, the restoring force for protein concentrations is small.

If we were to consider the scaled relaxation rates $\tilde{\lambda} = \frac{\lambda}{b_{0,3}}$, where $b_{0,3} = K \tau_m$ is the relaxation rate for proteins when there are no transcriptional regulation (which is also the growth rate per unit volume $\mu_3$ in the limit $\tau_p \to \infty$), then
\begin{equation}
    \tilde{\lambda} = \frac{1}{2}\left(-\omega_3 \pm \sqrt{\omega_3^2 + 4 \lambda_M (\omega_3+1)} \right) - 1,
    \label{eqn:lambdatilde_phase3}
\end{equation}
where $\omega_3 = \frac{1}{K\tau_m^2}-1$. This expression is the same as that in phases 1 and 2 (Eqns.\ref{eqn:lambda_phase1},\ref{eqn:lambda_phase2}), with $\omega$ and $\omega_2$ now replaced by $\omega_3$. Therefore, as before, the system is always stable as long as $\lambda_M < 1$ and unstable if $\lambda_M > 1$, regardless of the value of $\omega_3$.

\section{Effect of sign of regulatory interactions on stability}
\label{sec:IntSignEffect}

In this section, we investigate how the relative fraction of up- and down- regulatory interactions affect the maximal eigenvalue $\lambda_{M,r_{max}}$ of the interaction matrix. 

\begin{figure}[!htb]
    \centering
	\includegraphics[width=8.7cm]{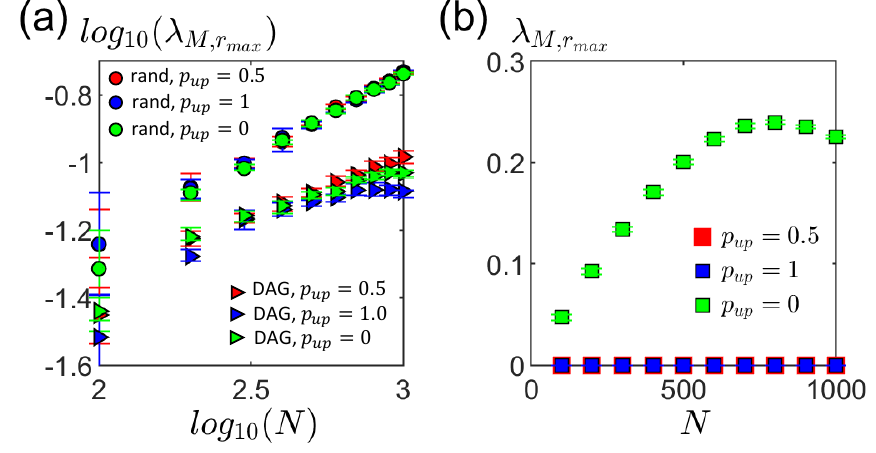}
    \caption{Effect of sign of regulatory interactions on maximum eigenvalue $\lambda_{M,r_{max}}$. (a) For fully random (circles) and random DAG (triangles) regulatory networks, $\lambda_{M,r_{max}}$ is approximately the same when there is an equal fraction of up- and down- regulatory interactions (red markers, '$p_{up} = 0.5$'), when all interactions are up-regulating (blue markers, '$p_{up} = 1$') and when all interactions are down-regulating (green markers, '$p_{up} = 0$'). [Parameters: $k_p = 1$, $\rho = 0.01$, $n = 1$, $\Omega_{max} = 1.5$.] (b) For bipartite interaction networks, up- regulating interactions are stabilizing while down-regulating interactions are destabilizing. These results hold in all phases of the model (since it is always the same interaction matrix that matters), even though the physical origin of the background, indirect interactions change between the three phases. [Parameters: number of TFs $q = 0.1N$, other parameters same as in (a).] }
    \label{SIfig:EffectofIntSign}
\end{figure}

We find that for random and DAG networks, the fraction of up-regulating interactions $p_{up}$ does not significantly affect $\lambda_{M,r_{max}}$ (Fig. \ref{SIfig:EffectofIntSign}a). However, for bipartite networks (which do not have any direct interactions between TFs), having only down- regulating interactions ($p_{up} = 0$) increases $\lambda_{M,r_{max}}$ dramatically compared to the scenario of having $p_{up} = 0.5$ (Fig. \ref{SIfig:EffectofIntSign}b). This is consistent with the tendency for inhibitory (activating) regulations to destabilize (stabilize) the system, which comes from the indirect effect that a regulator has on itself: a slight increase in the concentration of an inhibitor from its steady-state value will reduce the gene copy number and hence mRNA levels of the regulated gene. The mRNAs of the inhibitor therefore make up a larger fraction of the total mRNA in the cell. When ribosomes are limiting (phases 1 and 2), all mRNAs compete for the shared pool of ribosomes, and a higher mRNA fraction therefore causes the inhibitor concentrations to increase further. In phase 3, the reduction in mRNA levels of the regulated gene reduces the rate at which proteins are made. This slowing down of the increase in cell volume causes the inhibitor protein concentration to increase. This effect is much smaller in the case of random and DAG networks because their stability is dominated by the stronger, direct interactions among TFs.

\section{Effect of distribution of fold-change $\Omega_{ij}$ on stability}
\label{sec:FCdistEffect}

In this section, we investigate the effect that the distribution of fold-changes $\Omega_{ij}$ of the regulatory interactions has on the maximal eigenvalue $\lambda_{M,r_{max}}$ of the interaction matrix.

\begin{figure}[!htb]
    \centering
	\includegraphics[width=8.7cm]{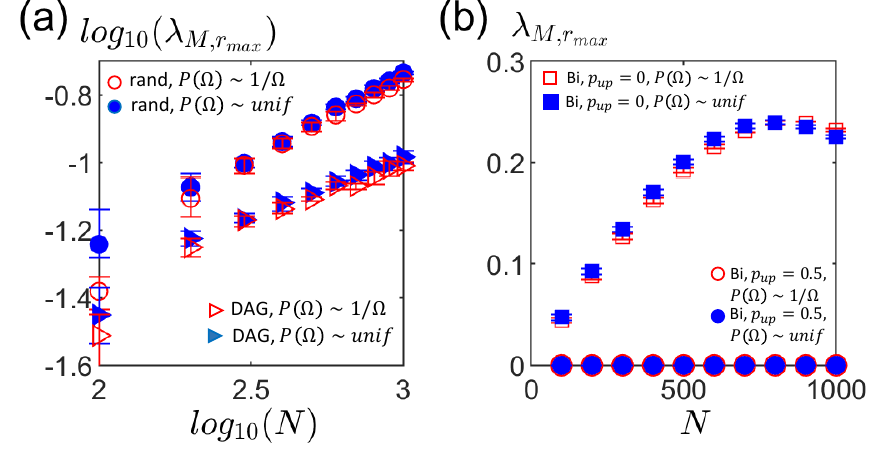}
    \caption{Effect of fold change distribution on how the maximal eigenvalue $\lambda_{M,r_{max}}$ varies with $N$. (a) For fully random (circles) and random DAG (triangles) regulatory networks, $\lambda_{M,r_{max}} \sim \sqrt{N}$ both when fold-changes $\Omega_{ij}$ are drawn from a uniform distribution (blue markers), and when they are drawn from a $1/\Omega$ distribution (red markers). [Parameters: $k_p = 1$, $\rho = 0.01$, $n = 1$, $\Omega_{max} = 1.5$, $p_{up} = 0.5$] (b) For bipartite interaction networks, the qualitative behavior of how $\lambda_{M,r_{max}}$ varies with $N$ is also independent of $P(\Omega)$ regardless of the fraction of up-regulating interactions $p_{up}$ [Parameters: number of TFs $q = 0.1N$, other parameters same as in (a).] }
    \label{SIfig:EffectofFCdist}
\end{figure}

In the main text, all the simulations were carried out with $\Omega$ drawn from a uniform distribution. For any fixed value of $\Omega_{max}$, having $P(\Omega) \sim \frac{1}{\Omega}$ (such that the \textit{logarithm} of $\Omega$ is uniformly distributed \cite{berger2015introduction, fewster2009simple}) would result in a lower $\langle \Omega \rangle$ and a higher fraction of weaker interactions. Nevertheless, we find that the qualitative behavior of how $\lambda_{M,r_{max}}$ scales with $N$ remains unchanged (Fig. \ref{SIfig:EffectofFCdist}). 
% We therefore expect a lower $\lambda_{M,r_{max}}$.

\section{Bipartite regulatory network}
\label{sec:Bipartite}

\subsection{Eigenvalue of Jacobian matrix}
For a bipartite regulatory network, the relevant $q \times q$ sector of the Jacobian matrix $\mathbf{Q}$ (main text Eqn. \ref{eqn:Qmatdef}) is given by:
\begin{equation}
    \mathbf{Q} = -\vec{c} \vec{a}^T,
\end{equation}
where $c_i$ is the concentration of TF $i$, and $a_i = \frac{\partial log g_T}{\partial c_i}$. Since this is a rank-1 matrix, it only has one eigenvalue $\lambda_b$ with corresponding eigenvector $\vec{v}$ such that
\begin{equation}
    -c_i \sum_j a_j v_j = \lambda_b v_i,
\end{equation}
for all $i = 1,2,...,q$. This implies that
\begin{equation}
     -\sum_i a_i c_i \sum_j a_j v_j = \lambda_b \sum_i a_i v_i.
\end{equation}
Therefore
\begin{equation}
     \lambda_b = -\sum_i a_i c_i,
\end{equation}
and
\begin{equation}
     v_i = c_i.
\end{equation}

\section{Effect of density of TF-otherTF interactions $\rho_q$ on maximum eigenvalue $\lambda_{M,r_{max}}$}
\label{sec:rhoqEffect}

Here, we investigate how the density of TF-otherTF interactions $\rho_q$ affects the maximal eigenvalue $\lambda_{M,r_{max}}$ of the interaction matrix. We find that without any auto-regulation loops, increasing $\rho_q$ increases $\lambda_{M,r_{max}}$, which is consistent with our observation that the probability of the system going unstable increases when $\rho_q$ is too large (Fig. \ref{fig:VaryTTint}b). 

\begin{figure}[!ht]
    \centering
	\includegraphics[width=4.45cm]{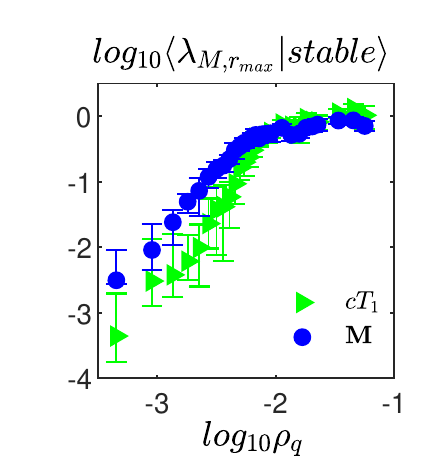}
    \caption{Effect of density $\rho_q$ of TF-otherTF interactions on the maximal eigenvalue of the interaction matrix $\lambda_{M,r_{max}}$. When the number of self-interactions $u_s = 0$ and the number of TF-otherTF interactions $u_n = 5000$, the average maximum eigenvalue $\lambda_{M,r_{max}}$ of the interaction matrix (blue circles) among stable systems increases with the density $\rho_q$ of TF-otherTF interactions, until reaching a threshold value of $\rho_q$ above which where it stays approximately constant. The maximum eigenvalue of the corresponding matrix $\mathbf{Q_1}$ consisting only of the direct interactions i.e. $Q_{1,ij} = c_i \frac{\partial \text{log}g_i}{\partial c_j}$ (green triangles) can be lower when $\rho_q$ is small. (other parameters: $N = 2274$, $q = 211$, $k_p = 1$, $n=2$, $\Omega_{max} = 1000$.)}
    \label{SIfig:VaryTTint}
\end{figure}

These values of $\lambda_{M,r_{max}}$ can be higher than the maximum eigenvalue of the corresponding matrix $\mathbf{Q_1}$ consisting only of the direct interactions i.e. $Q_{1,ij} = c_i \frac{\partial \text{log}g_i}{\partial c_j}$ (Fig. \ref{SIfig:VaryTTint}), especially for small values of $\rho_q$, suggesting that the indirect interactions can potentially play a role in affecting the stability of the system. In fact, in the limit where $\rho_q = 0$ (i.e. bipartite network), stability is only determined by these indirect interactions. Nevertheless, these indirect interactions are much weaker than the direct interactions, which accounts for the stability of the system at low $\rho_q$.

%unlike in RMT, $\lambda_{M,r_{max}}$ does not scale as $\sqrt{N}$ (Fig. \ref{SIfig:VaryTTint}). 
%In the presence of a large number of auto-regulatory interactions, the relaxation rate of the system is significantly reduced, and in this case, $\lambda_{M,r_{max}}$ is approximately independent of $\rho_q$.

%\caption{Effect of density $\rho_q$ of TF-otherTF interactions on the maximal eigenvalue of the interaction matrix $\lambda_{M,r_{max}}$. When the number of self-interactions $u_s = 0$ (blue circles), the average maximum eigenvalue $\tilde{\lambda}_{M,r_{max}}$ of the interaction matrix (among stable systems) increases with the density $\rho_q$ of TF-otherTF interactions, until reaching a threshold value of $\rho_q$ above which where it stays approximately constant. When $u_s = 134$ (red and green circles), $\tilde{\lambda}_{M,r_{max}}$ is higher at low values of $\rho_q$ and stays approximately constant throughout the range of $\rho_q$. (other parameters: $N = 2274$, $q = 211$, $k_p = 1$, $n=2$, $\Omega_{max} = 1000$.)}

\section{Allowing post-translational modifications in the model}
\label{sec:PTM}

Suppose each protein has to undergo some form of modification before they can be functional. In phase 1, the dynamics of the number of mRNAs $m_i$, proteins $p_i$, functional proteins $p^*_i$ of each gene $i$ can in general be written as:

\begin{align}
    \frac{dm_i}{dt} &= k_m \phi_i(\vec{c^*})n^* - \frac{m_i}{\tau_m} \\
    \frac{dp_i}{dt} &= k_p \frac{m_i}{\sum_j m_j} r^* - \frac{p_i}{\tau_p} - \psi_i(\vec{c^*}) p_i \\
    \frac{dp^*_i}{dt} &= \psi_i(\vec{c^*}) p_i - \frac{p^*_i}{\tau_p}, 
    \label{eqn:dynamicsPTM_numbers}
\end{align}
where $\phi_i$ is the gene allocation fraction as defined in our main model (which now depends on concentrations of the functional forms of the transcription factors), and we have assumed that the rate of modification of $p_i$ into $p^*_i$ is proportional to $p_i$, which is analogous to the reaction being first order in the substrate $i$. The modification rate also depends on the concentrations of other proteins involved in the modification through the function $\psi_i(\vec{c^*})$. As an example, for the case of a 1-step enzymatic reaction e.g. phosphorylation of a protein $i$ by a kinase $K$: $i + K \underset{k_b}{\stackrel{k_f}{\rightleftharpoons}} i\mhyphen K \xrightarrow{k_c} i^* + K$, solving the dynamics of each component ($i$, $i^*$, $K$ and $i\mhyphen K$) leads to $\psi_i(\vec{c^*}) = \frac{k_f k_c}{k_b+k_c} c^*_K$, where $c^*_K$ is the concentration of the \textit{free} kinase, and we have made the approximation that the concentration of the intermediate complex $i\mhyphen K$ does not change on the time-scale of the formation of product $i^*$ (quasi-steady-state approximation). If we further assume that the total kinase concentration (including both the free and bound forms) is fixed at a constant value of $c^*_{K,tot}$, we recover the familiar Michaelis-Menten kinetics, with $\psi_i(\vec{c^*}) = \frac{k_c c^*_{K,tot}}{k_m + c_i}$, where $c_i$ is the concentration of $i$ and $k_m = \frac{k_b + k_c}{k_f}$.

Approximating the volume of the cell as $V = \sum_i p_i + p^*_i$, the dynamics of concentrations are then given by:
\begin{align}
    \frac{dc_{mi}}{dt} &= k_m \phi_i(\vec{c^*})c_n^* - c_{mi} \left(k_p c^*_r + \frac{1}{\tau_m} - \frac{1}{\tau_p} \right)   \\
    \frac{dc_i}{dt} &= k_p c^*_r \left( \frac{c_{mi}}{\sum_j c_{mj}} - c_i \right) -  \psi_i(\vec{c^*}) c_i \\
    \frac{dc^*_i}{dt} &= \psi_i(\vec{c^*}) c_i - k_p c^*_r c^*_i. 
    \label{eqn:dynamicsPTM_concs}
\end{align}

With such a model, it is then possible to investigate how the coupling of both the transcriptional regulatory network and the network of post-translational modifications would affect stability of the system. We leave this interesting question for future work. 
    
%\bibliographystyle{ieeetr}
%\bibliography{SIRef}

\end{document}